\documentclass[11pt]{article}
\usepackage[T1]{fontenc}
\usepackage{amssymb,amsmath,amsfonts,dsfont,amsthm,bm}
\usepackage{float,rotfloat} 
\usepackage{mathrsfs}
\usepackage{multirow}
\usepackage{enumerate}
\usepackage{enumitem}
\usepackage{mathtools}
\usepackage{hyperref,url}
\usepackage{graphicx}
\usepackage[hang,small,bf]{caption}
\usepackage[top=1in, bottom=1in, left=1.0in, right=1.0 in]{geometry}
\usepackage{longtable}
\usepackage{color,xcolor}
\usepackage[round,sort,comma]{natbib}

\usepackage[title]{appendix}
\usepackage{etoolbox} 
\usepackage[english]{babel}
\usepackage[autostyle, english = american]{csquotes}
\MakeOuterQuote{"}

\patchcmd{\appendices}{\quad}{: }{}{} 

\newcommand{\ID}{\mathds{1}}
\newtheorem{thm}{Theorem}
\newtheorem{prop}{Proposition}
\newtheorem{lemma}{Lemma}
\newtheorem{defn}{Definition}
\numberwithin{thm}{section}
\numberwithin{prop}{section}
\numberwithin{lemma}{section}
\numberwithin{defn}{section}
\numberwithin{equation}{section}
\numberwithin{figure}{section}
\numberwithin{table}{section}
\newtheorem{note}{Note}[section]
\newcommand{\E}{\mathbb{E}}
\newcommand{\R}{\mathbb{R}}
\DeclareMathOperator{\csch}{csch}

\newcommand{\blue}{\color[rgb]{0,0,1}}
\definecolor{darkblue}{rgb}{0,0,0.55}

\allowdisplaybreaks

\hypersetup{,colorlinks=true,allcolors=darkblue}

\begin{document}
	
	\baselineskip 5mm
	
	\thispagestyle{empty}
	
	\begin{center}
		{\LARGE 
			Truncated, Censored, and Actuarial 
			Payment--type Moments for \\[10pt]
			Robust Fitting of a Single-parameter 
			Pareto Distribution
		}
		
		\vspace{15mm}
		
		{\large
			Chudamani Poudyal, Ph.D. \\
			{\sc\blue E-mail\/}: 
			\url{chudapsw@gmail.com}
		}
		
		\vspace{10mm}
		
		\copyright \
		Copyright of this Manuscript is held by the Author!
	\end{center}
	
	\vspace{3mm}
	
	\begin{quote}
		{\bf\em Abstract}.
		With some regularity conditions maximum likelihood estimators (MLEs)
		always produce asymptotically optimal (in the sense of
		consistency, efficiency, sufficiency, and unbiasedness) estimators. 
		But in general, the MLEs lead to non-robust statistical
		inference, for example, pricing models and risk measures.
		Actuarial claim severity is continuous, right-skewed, 
		and frequently heavy-tailed. 
		The data sets that such models are usually fitted to contain
		outliers that are difficult to identify and separate from genuine data. 
		Moreover, due to commonly used actuarial ``loss control strategies'' 
		in financial and insurance industries,
		the random variables we observe and wish to model are affected by truncation 
		(due to deductibles), censoring (due to policy limits), scaling 
		(due to coinsurance proportions) and other transformations.
		To alleviate the lack of robustness of MLE-based inference in risk 
		modeling, here in this paper, we propose and develop a new method
		of estimation -- {\em method of truncated moments\/} (MTuM)
		and generalize it for different scenarios of  
		loss control mechanism.
		Various asymptotic properties of those 
		estimates are established by using central limit theory. 
		New connections between different estimators are found. 
		A comparative study of newly-designed methods with 
		the corresponding MLEs is performed. 
		Detail investigation has been done for a single 
		parameter Pareto loss model including a simulation study.
		
		\vspace{4mm}
		
		{\bf\em Keywords \& Phrases\/}. 
		Claim Severity;
		Deductible;
		Relative Efficiency;
		Loss Models; 
		Robust Estimation; 
		Truncated and Censored Moments.
	\end{quote}
	
	\newpage
	
	\baselineskip 8mm
	
	\setcounter{page}{2}
	
	\section{Introduction}
	\label{sec:Introduction}
	
	The research leading to the results of this 
	work is basically motivated to find some
	trade-offs between robustness and efficiency
	of parametric estimators for ground-up 
	continuous loss distributions. 
	Parametric statistical loss models for insurance claim 
	severity are continuous, right-skewed, and frequently 
	heavy-tailed \cite{MR3890025}.
	The data sets that such models are usually fitted to contain 
	outliers that are difficult to identify and separate from genuine data. 
	As a result, there could be a significant difference in 
	statistical inference if there is a small perturbation 
	in the assumed model from the unknown true underlying 
	parametric model.
	In practice, due to commonly used loss control mechanism in the 
	financial and insurance industries \cite{EPUS}, 
	the random variables we observe and wish to model are affected 
	by data truncation (due to deductibles), censoring 
	(due to policy limits), and scaling (due to coinsurance factor).
	Maximum likelihood estimators (MLEs) typically result 
	in sensitive loss severity models
	if there is a small perturbation in the underlying assumed model
	or if the observed sample is coming from a contaminated distribution
	\cite{MR0120720}. 
	The implementation of MLE procedures even on ground-up loss 
	data is computationally challenging 
	\cite{MR3765327,MR3765333}.
	This issue is even more evident when one tries to fit complicated 
	multi-parameter models such as mixtures of Erlangs 
	\citep{REYNKENS201765,vgabl15}. 
	Thus, beside many ideas from the mainstream robust 
	statistics literature
	\cite[see, e.g.,][]{MR0362657,MR0161415,MR2488795},
	actuaries have to deal with 
	heavy-tailed and skewed distributions, data truncation 
	and censoring, identification and recycling of outliers,
	and aggregate loss, etc. 
	Based on a general class of $L$--statistics \citep[][]{MR0203874},
	two board classes of robust estimator  -- the methods of 
	{\em trimmed moments\/} (MTM) \cite{MR2497558}; 
	and {\em winsorized moments\/} (MWM) \cite{zbg18a}
	are recently developed with actuarial applications in view. 
	Therefore, it is appealing to search some estimation procedures
	which directly work with those mentioned loss control mechanism
	and are insensitive.
	
\begin{figure}[hbt!]
		\begin{flushleft}
			\makebox[\textwidth]
			{\includegraphics[width=0.98\textwidth]{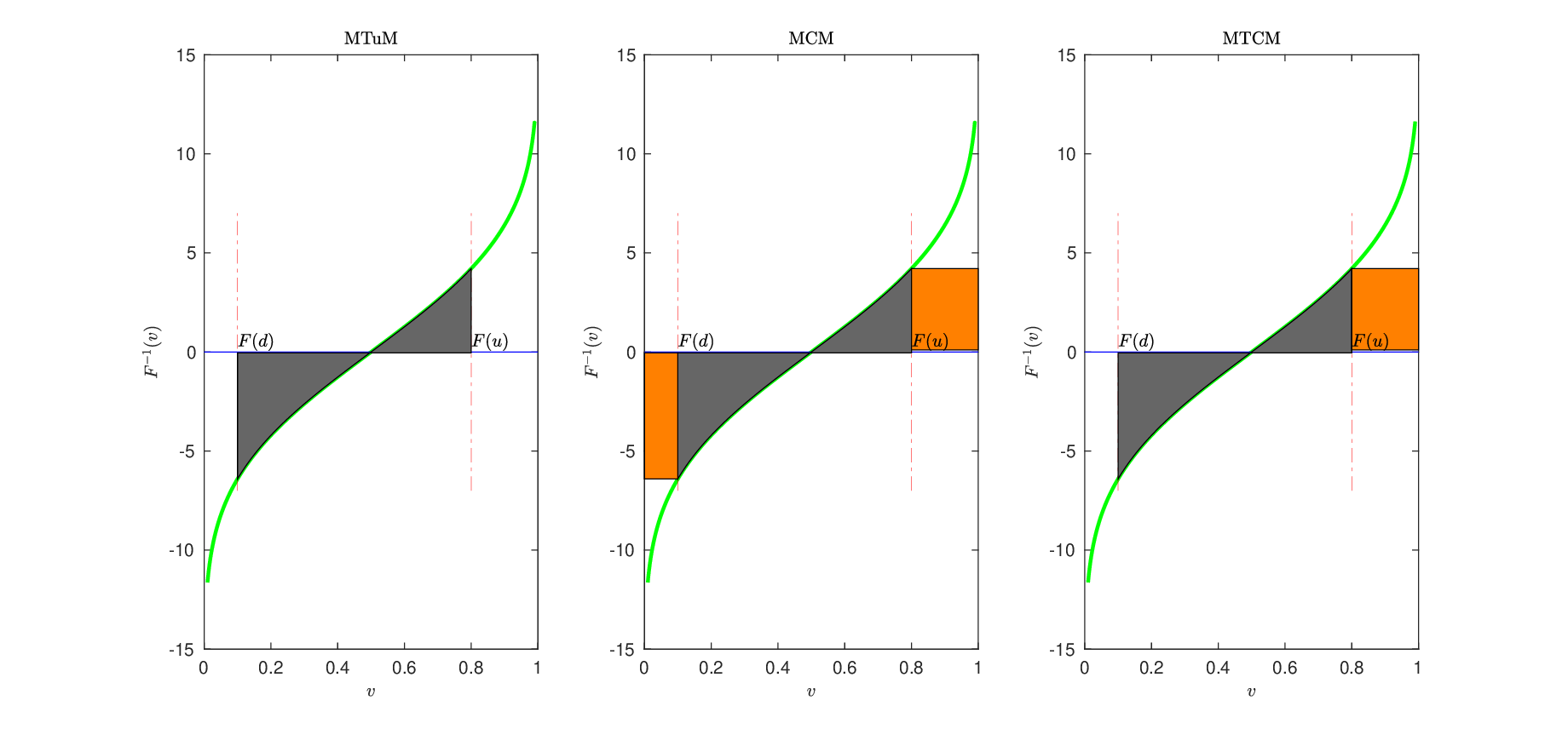}}
			\vspace{-0.8cm}
			\begin{table}[H]
				\centering
				{\scriptsize
					\begin{tabular}{lcl}
						MTuM & - & Method of Truncated Moments \\
						MCM & - & Method of Censored Moments \\
						MTCM & - & Method of Left Truncated and Right Censored Moments
					\end{tabular}
				}
			\end{table}
			\vspace{-0.5cm}
			\caption{
				Effects of MTuM (left panel), MCM (middle panel), and MTCM (right panel)
				on the underlying quantile function and thus data.
				MTuM focuses on the data only between
				the truncation thresholds, 
				MCM is a threshold censored form that
				takes into account the upper and lower outside values 
				as well (orange area), 
				and MTCM is a mixed version of both
				MTuM (left truncated) and MCM (right censored).}
			\label{fig:MTuMCases}
		\end{flushleft}
	\end{figure}

	If a truncated (both singly and doubly) normal sample data is
	available then the MLE procedures for such data have been developed 
	by \cite{MR0038041} and the method of truncated moments 
	estimators can be found in \cite{MR0045361} and \cite{MR0196848}.
	But the goal and motivation of this research work is different and 
	is initially purposed 
	by this author in \cite{MR3864903}.
	That is, instead of truncated sample data, we assume
	a complete ground-up sample loss data is available,
	i.e., the data set is neither truncated nor censored,
	and we propose and develop robust estimation procedures
	for the corresponding ground-up loss severity models.
	Instead of trimming or winsorizing a fixed lower (say, 2\%)
	and upper proportion (say, 3\%) of the observed sample data, 
	in this paper we develop a novel fixed lower and upper 
	thresholds method of {\em truncated moments\/} (say, MTuM)
	approach where the tail probabilities will be random. 
	Depends on the nature of the loss data mentioned above, 
	some variants of MTuM, called methods of fixed {\em censored moment\/} 
	(MCM) and {\em actuarial payment-type moment} (MTCM) will be defined 
	for single parameter Pareto distribution, 
	see Figure \ref{fig:MTuMCases}.
	Asymptotic distributions, such as normality and consistency,
	along with asymptotic relative efficiency of 
	those estimators with respect to the corresponding MLEs 
	are established. 
	Several theoretical connections between different 
	approaches are also discovered. 
	The newly designed procedures work like the standard 
	method-of-moments but instead of classical moments
	they are truncated or censored moments for a completely
	observed sample. 
	Irrespective with the heaviness of the underlying 
	distribution, threshold truncated and censored
	moments are always finite. 
	
	The remainder of the paper is organized as follows. 
	In Section \ref{sec:MTuM}, the newly proposed MTuM estimation
	procedure is defined in general with the establishment 
	of the corresponding asymptotic distributional properties. 
	In Section \ref{sec:MTuM_ExpPareto}, 
	we develop specific formulas of different
	estimators (including MTuM) when the underlying loss 
	distribution is Pareto I which is equivalent to an
	exponential distribution, and compare the asymptotic
	relative efficiency of all the estimators with respect to 
	the corresponding MLEs for completely observed data.
	Several connections among different estimators are established.
	Section \ref{sec:SimStudy} summarizes a detail simulation
	study of different estimators developed in this paper.
	Concluding remarks are offered in Section \ref{sec:Conclusion}.
	Finally, some additional results are provided in 
	Appendix \ref{apdx:allScenarios} and \ref{apdx:proofs}.
	
	\section{Method of Truncated Moments}
	\label{sec:MTuM}
	
	We assume that a complete ground-up loss data is available,
	i.e., the data set is neither truncated nor censored.
	Then, instead of trimming or winsorizing fixed proportion from both 
	tails, from a completely observed data, as investigated by
	\cite{MR2497558,zbg18a}, in this approach of parametric 
	estimation we truncate the data from below at lower threshold
	and from above at upper threshold and then apply the method 
	of moments on the remaining data. 
	We call such an approach {\em method-of-truncated-moments 
		(MTuM -- for short)}. 
	
	\subsection{Definition}
	\label{sec:MTuMDef}
	
	Let $X_{1}, X_{2},..., X_{n}$ be {\em i.i.d.} random variables
	with common ground-up cdf $F(\cdot|\bm{\theta})$, where
	$\bm{\theta} := (\theta_{1}, \ldots, \theta_{k}), \ k \geq 1$ 
	is the parameter vector to be estimated. The truncated moments
	estimators of $\theta_{1}, \theta_{2},...,\theta_{k}$ are 
	computed according to the following procedures.
	
	\begin{enumerate}[label=(\roman*)]
		\item 
		The sample truncated moments are computed as 
		\begin{equation} \label{eq:sample_mtum}
			\widehat{\mu}_{j} = \frac{\sum_{i=1}^{n} h_{j}(X_{i})
				\ID \{d_{j} < X_{i} \leq u_{j} \}}{\sum_{i=1}^{n} 
				\ID \{d_{j} < X_{i} \leq u_{j} \}},\ \ \ \ \ 1 \leq j \leq k,
		\end{equation} 
		where $\ID \{\cdot\}$ denotes the indicator function. 
		The $h_{j}'s$ in (\ref{eq:sample_mtum}) are specially
		chosen functions as well 
		as the thresholds $d_{j}$ and $u_{j}$ are chosen by the researcher. 
		In general, 
		it is reasonable to assume that 
		$X_{1:n}\leq d_{j} < u_{j} \leq X_{n:n}$, for 
		all $1 \leq j \leq k$, where $X_{1:n}$ and $X_{n:n}$ 
		are the smallest and the 
		largest order statistics, respectively, from the sample.
		
		\item
		Derive the corresponding population truncated moments as
		\begin{align} 
			\label{eq:pop_mtum} 
			\mu_{j}
			\left(
			\theta_{1}, \theta_{2},...,\theta_{k}
			\right) 
			=  
			\mathbb{E}
			\left[
			h_{j}(X)|d_{j}<X \leq u_{j}
			\right]
			& = 
			\frac{\mathbb{E}
				\left[
				h_{j}(X)\ID \{d_{j} < X \leq u_{j} \}
				\right]}
			{\mathbb{P}(d_{j}<X \leq u_{j})} \nonumber \\
			& = 
			\frac{\int_{d_{j}}^{u_{j}}h_{j}(x)f(x|\bm{\theta}) \, dx}{F(u_{j}|\bm{\theta})-F(d_{j}|\bm{\theta})},\ \ \ 1\leq j \leq k.
		\end{align} 
		
		\item 
		Now, match the sample and population truncated moments from 
		(\ref{eq:sample_mtum}) and (\ref{eq:pop_mtum}) to get the following 
		system of equations for $\theta_{1},\theta_{2},...,\theta_{k}:$
		\begin{equation} 
			\label{eq:match_mtum}
			\left\{
			\begin{array}{lcl}
				\mu_1 (\theta_1, \ldots, \theta_k) & = & \widehat{\mu}_1 \\
				& \vdots & \\
				\mu_k (\theta_1, \ldots, \theta_k) & = & \widehat{\mu}_k  \\
			\end{array} \right.
		\end{equation}
	\end{enumerate}
	
	\begin{defn} 
		A solution to the system of equations
		(\ref{eq:match_mtum}), say 
		$
		\widehat{\bm{\theta}}
		=
		\left(
		\widehat{\theta}_{1},
		\widehat{\theta}_{2},...,
		\widehat{\theta}_{k}
		\right),
		$
		if it exists, is called the 
		{\textit{method of truncated moments (MTuM)}} 
		estimator of $\bm{\theta}$.
		Thus, 
		$\widehat{\theta}_{j}
		=:
		g_{j}
		\left(
		\widehat{\mu}_{1},\widehat{\mu}_{2},..., \widehat{\mu}_{k}
		\right)$,
		$1\leq{j}\leq{k}$ 
		are the MTuM estimators of $\theta_{1},\theta_{2},...,\theta_{k}$.
	\end{defn}
	
	\begin{note}
		Obviously, it is possible that the system of 
		equations (\ref{eq:match_mtum}) 
		does not have a solution, or it is difficult to
		solve the
		system even with numerical
		methods when $k$ is large. 
		To facilitate this issue, the functions $h_{j}$ 
		have to be chosen carefully. But most claim severity distributions have 
		a small number $k$ of parameters, usually not
		exceeding three 
		\citep[see][Appendix A]{MR3890025}. 
		\qed
	\end{note}
	
	\subsection{Asymptotic Properties}
	\label{sec:mtum_asymptotic_properties}
	
	For $1 \leq j,j' \leq k$ and for any positive integer $n$, 
	define 
	$\ID\{d_{jj'}<X \leq u_{jj'}\} := \ID\{d_{j}<X \leq u_{j}\}\ID\{d_{j'}<X \leq u_{j'}\}$
	and consider the following additional notations: 
	\begin{align*}
		Z_{j} & := h_{j}(X), & h_{jj'}(x) & := h_{j}(x)h_{j'}(x), 
		& p_{j} & := F(u_{j}|\bm{\theta}) - F(d_{j}|\bm{\theta}), \\
		Y_{jj'} & := Y_{j}Y_{j'}, &  Y_{j} & := Z_{j}\ID\{d_{j}<X\leq u_{j}\}, & p_{jj'} 
		& := F(u_{jj'}|\bm{\theta}) - F(d_{jj'}|\bm{\theta}), \\
		r_{j} & := h_{j}(d_{j}), 
		\ R_{j} := h_{j}(u_{j}), & W_{jj'} & := Z_{j}\ID\{d_{jj'}<X\leq u_{jj'}\}, 
		& p_{j,n} & := F_{n}(u_{j}) - F_{n}(d_{j}), 
	\end{align*}
	where $F_{n}(x) = \frac{1}{n}\sum_{i=1}^{n}\ID\{X_{i} \leq x\}$ is the empirical 
	distribution function.
	Note that $Y_{jj'} = Y_{j'j}$ but $W_{jj'} \neq W_{j'j}$ for $j\neq j'$, 
	in general. With those notations, the density of $Y_{j}, \ (1\leq j \leq k)$ 
	can be expressed as
	\begin{align*}
		f_{Y_{j}}(y) & = 
		\begin{cases}
			1-F_{Z_{j}}(R_{j}|\bm{\theta})+F_{Z_{j}}(r_{j}|\bm{\theta}), 
			& \mbox{if } y = 0;\\
			f_{Z_{j}}(y|\bm{\theta}), 
			& \mbox{if } r_{j} < y < R_{j}; \\
			0, & \mbox{otherwise.}
		\end{cases} 
	\end{align*}
	The density of the random variables $Y_{jj'} = Y_{j'j}$ and $W_{jj'}$ 
	can be constructed with the four possible scenarios which are listed in 
	Appendix \ref{apdx:allScenarios}. 
	To establish the asymptotic distribution of 
	$\widehat{\bm{\mu}}$, we need the following lemma.
	
	\begin{lemma} 
		\label{lemma:YpCov}
		For $1\leq j,j' \leq k$,
		\begin{alignat*}{6}
			\pushQED{\qed} 
			\mathbb{C}{ov}
			\left(
			Y_{j},Y_{j'}
			\right) 
			& = 
			\mu_{\mbox{\tiny $Y_{jj'}$}}
			-
			\mu_{\mbox{\tiny $Y_{j}$}}
			\mu_{\mbox{\tiny $Y_{j'}$}}, \
			& 
			\mathbb{C}{ov}
			\left(
			Y_{j};p_{j',1}
			\right)
			& =
			\mu_{\mbox{\tiny $W_{jj'}$}}
			-
			\mu_{\mbox{\tiny $Y_{j}$}} 
			p_{j'}, \ 
			& 
			\mathbb{C}{ov}
			\left(
			p_{j,1};p_{j',1}
			\right)  
			& = 
			p_{jj'} - p_{j}p_{j'}. 
			\tag*{\qedhere}
		\end{alignat*} 
	\end{lemma}
	
	Consider a $2k$ -- dimensional random vector
	$
	\bm{V}
	:=
	\left(
	Y_{1},\ldots,Y_{k},p_{1,1},
	\ldots,p_{k,1}
	\right).
	$ 
	Clearly the mean vector of $\bm{V}$ 
	is
	$
	\bm{\mu}_{\mbox{\tiny $\bm{V}$}}
	= 
	\left(
	\mu_{\mbox{\tiny $Y_{1}$}},
	\ldots,
	\mu_{\mbox{\tiny $Y_{k}$}},
	p_{1},
	\ldots,p_{k}
	\right)
	$ 
	and with Lemma \ref{lemma:YpCov},
	the variance-covariance 
	matrix is 
	$\bm{\Sigma}_{\bm{V}} 
	= 
	\left[\sigma_{\bm{V},j j'}^{2}\right]_{j,j'=1}^{2k}$, 
	where 
	\[
	\sigma_{\bm{V},j j'}^{2} = 
	\begin{cases}
	\mu_{\mbox{\tiny $Y_{j j'}$}}
	-
	\mu_{\mbox{\tiny $Y_{j}$}}\mu_{\mbox{\tiny $Y_{j'}$}},
	& \ \ \ \ \ 1 \leq j, j' \leq{k}; \\ 
	\mu_{\mbox{\tiny $W_{j(j'-k)}$}} 
	- 
	\mu_{\mbox{\tiny $Y_{j}$}}p_{\mbox{\tiny $j'-k$}}, 
	& \ \ \ \ \ 1 \leq j \leq{k}; k+1 \leq j' \leq 2k; \\
	\mu_{\mbox{\tiny $W_{(j-k)j'}$}} 
	-
	\mu_{\mbox{\tiny $Y_{j'}$}}p_{\mbox{\tiny $j-k$}}, 
	& \ \ \ \ \ 1 \leq j' \leq{k}; k+1 \leq j \leq 2k; \\
	p_{\mbox{\tiny $(j-k)(j'-k)$}} 
	-
	p_{\mbox{\tiny $j-k$}} p_{\mbox{\tiny $j'-k$}}, & 
	\ \ \ \ \ k+1 \leq j,j' \leq 2k.
	\end{cases} 
	\]
	
	\begin{thm} 
		\label{thm:asymptoticOfMTuMRatioVector}
		The empirical estimator 
		\begin{align*}
			\widehat{\bm{\mu}}_{V} 
			& := 
			\frac{1}{n}\left(\sum_{i=1}^{n}Y_{1,i},\ldots,\sum_{i=1}^{n}Y_{k,i},
			\sum_{i=1}^{n}p_{1,i},\ldots,\sum_{i=1}^{n}p_{k,i}\right) 
			= 
			\left(\overline{Y}_{1,n},\ldots,\overline{Y}_{k,n},p_{1,n},
			\ldots,p_{k,n}  \right)
		\end{align*}
		of the mean vector $\bm{\mu}_{V}$ is such that
		$
		\widehat{\bm{\mu}}_{V}
		\sim \mathcal{AN}\left(\bm{\mu}_{\bm{V}},\frac{1}{n}\bm{\Sigma}_{\bm{V}}\right).
		$
		\begin{proof} 
			Let $\{\bm{V}_{n}\}$ be a sequence of 
			$i.i.d.$ $\bm{V}$ random vectors, 
			then by multivariate Central Limit Theorem 
			\citep[see, e.g.,][Theorem B, p. 28]{MR595165}, 
			we have: 
			\[
			\left(\overline{Y}_{1,n},\ldots,
			\overline{Y}_{k,n},p_{1,n},\ldots,p_{k,n}  \right) 
			= 
			\frac{1}{n}\sum_{i=1}^{n}\bm{V_{i}} \sim \mathcal{AN}\left(\bm{\mu}_{\bm{V}},
			\frac{1}{n}\bm{\Sigma}_{\bm{V}}\right). \qedhere
			\]
		\end{proof}
	\end{thm}
	
	\noindent
	The system of MTuM equations (\ref{eq:match_mtum})
	can now be written as:
	\begin{align} 
		\label{eq:match_mtum2} 
		\left\{
		\begin{array}{lclcl}
			\mu_1 (\theta_1, \ldots, \theta_k) 
			& = &
			\widehat{\mu}_1 
			& = & 
			\frac{\overline{Y}_{1,n}}{p_{1,n}}, \\
			& \vdots & {} & \vdots  \\
			\mu_k (\theta_1, \ldots, \theta_k) 
			& = & 
			\widehat{\mu}_k  
			& = & 
			\frac{\overline{Y}_{k,n}}{p_{k,n}}.  \\
		\end{array} \right.
	\end{align}
	
	\begin{lemma} \label{lemma:differentialFun}
		Consider a function $g_{\bm{V}}:\R^{2k} 
		\rightarrow \R^k$ for $\bm{x} = (x_{1},x_{2},\ldots,x_{2k})$ 
		defined by 
		\[
		g_{\bm{V}}(\bm{x}) 
		= 
		\left(g_{1}(\bm{x}),\ldots,g_{k}(\bm{x})\right) 
		:= 
		\left(\frac{x_{1}}{x_{k+1}},\ldots, \frac{x_{k}}{x_{2k}}\right), 
		\]
		where $\ x_{i} \neq 0, \  i = k+1,\ldots, 2k$. 
		Then $g_{\bm{V}}$ is totally differentiable 
		at any point $\bm{x}_{0} \in \R^{2k}$. 
		\begin{proof}
			A proof directly follows from \cite[Lemma 1.12.2]{MR595165}.
		\end{proof}
	\end{lemma}
	
	With the help of Theorem \ref{thm:asymptoticOfMTuMRatioVector} and Lemma 
	\ref{lemma:differentialFun}, we are now ready to state the asymptotic 
	distribution of the truncated sample moment vector $\widehat{\bm\mu}$ 
	whose proof can be found in 
	Appendix \ref{apdx:proofs}.
	
	\begin{thm} 
		\label{thm:MTuMMuAsyDist}
		The asymptotic joint distribution of the truncated sample moment
		vector 
		$
		\left(\widehat{\mu}_{1},\ldots,\widehat{\mu}_{k}\right)
		$
		is given by 
		$
		N\left(\bm{\mu},\frac{1}{n}\bm{\Sigma}\right)
		$
		with 
		$
		\bm{\Sigma} 
		=
		{\bm D_{\bm{V}}\Sigma_{\bm{V}}D_{\bm{V}}'} 
		= :
		\left[\sigma_{jj'}^{2}\right]_{k\times k},
		$
		where
		\begin{align*}
			\sigma_{jj'}^{2} 
			& = 
			\frac{1}{p_{j'}} 
			\left(\frac{\mu_{\mbox{\tiny $Y_{jj'}$}}
				-
				\mu_{\mbox{\tiny $Y_{j}$}}
				\mu_{\mbox{\tiny $Y_{j'}$}}}
			{p_{j}}
			-
			\frac{\mu_{\mbox{\tiny $Y_{j}$}}
				\left(
				\mu_{\mbox{\tiny $W_{jj'}$}}
				-
				\mu_{\mbox{\tiny $Y_{j'}$}}
				p_{j}
				\right)
			}
			{p_{j}^{2}}
			\right)
			-
			\frac{\mu_{\mbox{\tiny $Y_{j'}$}}}
			{p_{j'}^{2}}
			\left(
			\frac{\mu_{\mbox{\tiny $W_{jj'}$}}
				-
				\mu_{\mbox{\tiny $Y_{j}$}}
				p_{j'}}{p_{j}}
			-
			\frac{\mu_{\mbox{\tiny $Y_{j}$}}
				\left(
				p_{jj'}-p_{j}p_{j'}
				\right)
			}
			{p_{j}^{2}}
			\right).
		\end{align*}
	\end{thm}
	
	Now, with 
	$\bm{\widehat{\mu}}
	=
	\left(\widehat{\mu}_{1},\ldots,\widehat{\mu}_{k} \right)$ 
	and
	$g_{\bm{\theta}}(\widehat{\bm{\mu}}) 
	= 
	\left(g_{1,\bm{\theta}}(\widehat{\bm{\mu}}),
	\ldots,g_{k,\bm{\theta}}(\widehat{\bm{\mu}}) \right)  
	= \widehat{\bm{\theta}}$,  
	then by delta method \citep[see, e.g.,][Theorem A, p. 122]{MR595165}, 
	we have the following main result of this section.
	\begin{thm} 
		\label{thm:mtumMuHatDist}
		The MTuM estimator of $\bm{\theta}$, denoted by
		$\widehat{\bm{\theta}}$, has the following asymptotic distribution: 
		\[
		\widehat{\bm{\theta}} 
		=
		\left(\widehat{\theta}_{1},\ldots,\widehat{\theta}_{k}\right) 
		\sim \mathcal{AN}\left(\bm{\theta},\frac{1}{n}{\bm{D\Sigma D'}}\right),
		\] 
		where the Jacobian $\bm{D}$ is given by
		$
		\bm{D}
		=
		\left[\left. \frac{\partial g_{j,\bm{\theta}}}{\partial \widehat{\mu}_{j'}}\right\vert_{\widehat{\bm{\mu}}
			=
			\bm{\mu}}\right]_{k\times k} =: \left[d_{jj'}\right]_{k\times k}
		$ and the variance-covariance matrix $\bm{\Sigma}$ has 
		the same form as in Theorem \ref{thm:MTuMMuAsyDist}.
	\end{thm}
	
	\begin{note} 
		\label{note:trimmed_is_truncated} 
		In view of the above derivations, we notice that data trimming and thus 
		(method of trimmed moments -- MTM) investigated by 
		\cite{MR2497558} can be interpreted as special cases of data truncation 
		and thus MTuM, respectively. To see that, let $F$ be the distribution 
		function of 
		$X$. For $1 \leq j \leq k$, consider $F(d_{j}|\bm{\theta})=a_{j}$ 
		and $F(u_{j}|\bm{\theta})=1-b_{j}$. 
		Then, using integration by substitution with $U=F(X)$, 
		the equation (\ref{eq:pop_mtum}) becomes
		\begin{subequations}
			\begin{align}
				\mu_{j}(\theta_{1}, \theta_{2},...,\theta_{k}) 
				= 
				\frac{\int_{d_{j}}^{u_{j}}h_{j}(x)f(x|\bm{\theta}) \, dx}{F(u_{j}|\bm{\theta})-F(d_{j}|\bm{\theta})} 
				& = 
				\frac{\int_{F(d_{j}|\bm{\theta})}^{F(u_{j}|\bm{\theta})}
					h_{j}(F^{-1}(u|\bm{\theta})) \, du}
				{F(u_{j}|\bm{\theta})-F(d_{j}|\bm{\theta})}
				\label{eq:mtm_mtum_mtum} \\
				& =
				\frac{\int_{a_{j}}^{1-b_{j}}h_{j}(F^{-1}(u|\bm{\theta})) \, du}
				{1-a_{j}-b_{j}}, \label{eq:mtm_mtum_mtm}
			\end{align}
		\end{subequations}
		which is equivalent to the corresponding population trimmed moment. 
		\qed
	\end{note}
	
	\begin{note} 
		For estimation purposes these two approaches (i.e., MTM and MTuM) are very 
		different. With the MTuM approach, the limits of integration as well as the 
		denominator in equation  (\ref{eq:mtm_mtum_mtum}) are unknowns, which create 
		technical complications when we want to assess the asymptotic properties of 
		MTuM estimators. 
		On the other hand, with the MTM approach, both the limits of 
		integration and the denominator in equation (\ref{eq:mtm_mtum_mtm}) are constants, 
		which simplify the matters significantly. Indeed, as is evident from complete 
		data examples in \cite{MR2497558} and 
		\cite{zbg18a}, MTM leads to explicit 
		formulas for all location-scale families and their variants, but that is not 
		the case with MTuM. In view of this, we will consider the MTuM approach further
		only for some data scenarios, but not all. 
		\qed
	\end{note}
	
	\section{Pareto I Distribution} 
	\label{sec:MTuM_ExpPareto}
	
	Let $Y \sim \text{Pareto I}(\alpha,x_{0})$ with the distribution function 
	$F_{Y}(y) = 1-\left(x_{0}/y\right)^{\alpha}, \ y > x_{0}$, 
	zero elsewhere, where $\alpha > 0$ is the shape (so called tail) 
	parameter and $x_{0} > 0$ is known left threshold.
	Then $X := log{\left(Y/x_{0}\right)} \sim \text{Exp}(\theta = 1/\alpha)$ 
	with the distribution function $F_{X}(x) = 1-e^{-x/\theta}$. 
	Therefore, in order to estimate $\alpha$ it is equivalent to
	estimate the exponential parameter $\theta$ that what we will proceed for the 
	rest of this section. MTuM will be derived with asymptotic results for a complete
	{\em i.i.d.} sample from an exponential distribution. 
	For this particular distribution, we also explore two additional methods: 
	method of censored moments and insurance payment--type  moment estimators. 
	Several connections between different approaches are established. 
	
	The asymptotic performance of the newly designed estimators will 
	be measured via asymptotic relative efficiency (ARE) with respect 
	to MLE and is defined as \citep[see, e.g.,][]{MR595165,MR1652247}:
	\begin{equation} 
		\label{eq:infinite_relative_efficiency_benchmark_MLE}
		ARE(\mathcal{C}, MLE) 
		= 
		\frac{\text{asymptotic variance of MLE estimator}}
		{\text{asymptotic variance of }\mathcal{C} \text{ estimator}}.
	\end{equation}
	The main reason why MLE should be used as a benchmark is its
	optimal asymptotic performance in terms of variability 
	(of course,  with the usual caveat of "under certain regularity conditions").
	
	\subsection{Method of Truncated Moments}
	\label{sec:MTuM_for_ExpPareto}
	
	In this section, we derive MTuM and related estimators for the parameter 
	of exponential distribution for completely observed data. Since there is 
	a single parameter, $\theta$, to be estimated, we consider the function 
	$h(x) = x$. 
	Let   $X_{1},\ldots,X_{n}$  be {\em i.i.d.\/} random variables given
	as in Definition \ref{sec:MTuMDef}. 
	Consider $d$ and $u$ be the left and right truncation points, 
	respectively. Then the sample truncated moment is given by
	\begin{align*}
		\widehat{\mu}_{\mbox{\tiny MTuM}}
		& := 
		\frac{
			\left( 
			\sum_{i=1}^{n} X_{i}\ID \{d < X_{i} \leq u\}
			\right)/n
		}
		{
			\left(
			\sum_{i=1}^{n} \ID \{d < X_{i} \leq u \}\right)/n}
		= 
		\frac{\left(\sum_{i=1}^{n}Y_{i}\right)/n}{F_{n}(u)-F_{n}(d)} 
		= 
		\frac{\overline{Y}_{n}}{F_{n}(u)-F_{n}(d)} 
		= 
		\frac{\overline{Y}_{n}}{p_{n}}, 
	\end{align*}
	where 
	$Y_{1},Y_{2},\ldots,Y_{n} \stackrel{i.i.d.}{\sim} Y 
	:= X\ID \{d < X \leq u\}$ and $p_{n} := F_{n}(u)-F_{n}(d)$ 
	with 
	$p
	\equiv 
	p(\theta)=F(u|{\theta}) -F(d|{\theta})
	=
	e^{-\frac{d}{\theta}}-e^{-\frac{u}{\theta}}$.
	
	\begin{thm} 
		\label{thm:theTruncatedMeanVariance}
		The mean and the variance of the random variable $Y$ are 
		respectively given by 
		\begin{align*}
			\mu_{Y} 
			& =  
			\theta p
			+ de^{-\frac{d}{\theta}} - ue^{-\frac{u}{\theta}} \quad \mbox{and} \quad
			\sigma_{Y}^{2} 
			= 
			2\theta^{2}\left(\Gamma\left(3;\frac{u}{\theta}\right) - \Gamma\left(3;\frac{d}{\theta}\right)\right) - \mu_{Y}^{2},
		\end{align*}
		where 
		$\Gamma(\alpha;x)$ with $\alpha>0, x>0$ is the incomplete
		gamma function defined as
		\begin{align*}
			\Gamma\left(\alpha;x\right) 
			& =
			\frac{1}{\Gamma\left(\alpha\right)}
			\int_{0}^{x}t^{\alpha -1}e^{-t} \, dt \quad \mbox{with} \quad 
			\Gamma(\alpha) 
			= 
			\int_{0}^{\infty}t^{\alpha -1}e^{-t} \, dt.
		\end{align*}
		\begin{proof}
			See Appendix \ref{apdx:proofs}.
		\end{proof}
	\end{thm}
	
	From Theorem \ref{thm:MTuMMuAsyDist}, 
	$\widehat{\mu}_{\mbox{\tiny MTuM}} 
	\sim 
	\mathcal{AN}\left(\frac{\mu_{Y}}{p},\frac{1}{n}
	\left( \frac{\sigma_{Y}^{2}}{p^{2}} 
	- \frac{(1-p)\mu_{Y}^{2}}{p^{3}}\right)\right)$. 
	Note that the asymptotic variance 
	of $\widehat{\mu}_{\mbox{\tiny MTuM}}$ 
	is exactly equal to the approximation through the second order 
	Taylor series expansion of the ratio of the asymptotic distribution of 
	$\overline{Y}_{n}$ and $p_{n}$ as mentioned in \cite{MRJHDANG}.
	The corresponding population version of 
	$\widehat{\mu}_{\mbox{\tiny MTuM}}$ is given by 
	\begin{eqnarray}
		\label{eqn:pop_MTuM1}
		\mu_{\mbox{\tiny MTuM}} 
		& := &
		\E[X|d<X \leq u] 
		= 
		\frac{\E[Y]}{F(u|{\theta})-F(d|{\theta})} 
		= 
		\frac{\mu_{Y}}{p}.
	\end{eqnarray}
	
	\begin{thm}
		\label{thm:ExpTruncatedMean}
		The equation 
		$\mu_{\mbox{\tiny MTuM}} 
		= 
		\widehat{\mu}_{\mbox{\tiny MTuM}}$ 
		has a unique solution $\widehat{\theta}$ 
		provided that 
		$d < \widehat{\mu}_{\mbox{\tiny MTuM}} < \frac{d+u}{2}$. 
		Otherwise, the solution does not exist.
		\begin{proof}
			It is clear that 
			$d < \widehat{\mu}_{\mbox{\tiny MTuM}} < u$. 
			Also, 
			$\mu_{\mbox{\tiny MTuM}}(\theta) 
			=
			\frac{\mu_{Y}}{p}
			= 
			\frac{e^{-\frac{d}{\theta}}\left(d+\theta\right) - e^{-\frac{u}{\theta}}\left(u+\theta\right)}{e^{-\frac{d}{\theta}} 
				- e^{-\frac{u}{\theta}}}$. 
			Then, in order to establish the result, it is enough 
			to prove the following statements:
			\[
			\begin{array}{lll}
			(a) \ \mu_{\mbox{\tiny MTuM}}(\theta) \ 
			\mbox{is strictly increasing,} &
			(b) \ 
			\displaystyle 
			\lim_{\theta \rightarrow 0+} 
			\mu_{\mbox{\tiny MTuM}}(\theta) 
			= d, \ \mbox{and} & 
			(c) \ 
			\displaystyle
			\lim_{\theta \rightarrow \infty} 
			\mu_{\mbox{\tiny MTuM}}(\theta) 
			= \frac{d+u}{2}.
			\end{array}
			\]
			First of all, let us establish that 
			$\mu_{\mbox{\tiny MTuM}}(\theta)$ is strictly increasing. 
			\begin{eqnarray*}
				\mu_{\mbox{\tiny MTuM}}^{'}(\theta) 
				& = & 
				\frac{\left(\left(\frac{d}{\theta}\right)^{2}e^{-\frac{d}{\theta}} 
					- \left(\frac{u}{\theta}\right)^{2}e^{-\frac{u}{\theta}}\right) 
					\left(e^{-\frac{d}{\theta}} - e^{-\frac{u}{\theta}} \right)
					+\left(e^{-\frac{d}{\theta}} - e^{-\frac{u}{\theta}} \right)^{2}}
				{\left(e^{-\frac{d}{\theta}} 
					- e^{-\frac{u}{\theta}} \right)^{2}} \\
				& = & 
				\frac{-e^{-\frac{d+u}{\theta}}\left(\left(\frac{u}{\theta}\right)^{2} 
					+ \left(\frac{d}{\theta}\right)^{2}\right)+\frac{2du}
					{\theta^{2}}e^{-\frac{d+u}{\theta}}+\left(e^{-\frac{d}{\theta}} 
					- e^{-\frac{u}{\theta}} \right)^{2}}{\left(e^{-\frac{d}{\theta}} 
					- e^{-\frac{u}{\theta}} \right)^{2}} \\
				& = & 
				1 - \frac{4\left(d-u\right)^{2}}{4\theta^{2}
					\left(e^{\frac{d-u}{2\theta}}
					- e^{-\frac{d-u}{2\theta}}\right)^{2}} 
				=
				1 - \left(\frac{d-u}{2\theta}\right)^{2} 
				\left(\frac{2}{e^{\frac{d-u}{2\theta}} 
					- e^{-\frac{d-u}{2\theta}}}\right)^{2} \\
				& = &
				1 - \left(\frac{d-u}{2\theta}\right)^{2} 
				\csch^{2}{\left(\frac{d-u}{2\theta}\right)}.
			\end{eqnarray*}
			Therefore, 
			$\mu_{\mbox{\tiny MTuM}}^{'}(\theta) > 0$ 
			if and only if $\left(\frac{d-u}{2\theta}\right)^{2} 
			< 
			\sinh^{2}{\left(\frac{d-u}{2\theta}\right)}$, 
			which is true since $x<\sinh{x}$ for all $x>0$ and 
			$x > \sinh{x}$ for all $x<0$. 
			Further,
			\begin{align*}
				\lim_{\theta \rightarrow 0+} 
				\mu_{\mbox{\tiny MTuM}}(\theta) 
				& = 
				\lim_{\theta \rightarrow 0+}\left[ \theta + \frac{de^{-\frac{d}{\theta}} - ue^{-\frac{u}{\theta}}}{e^{-\frac{d}{\theta}} - e^{-\frac{u}{\theta}}}\right] 
				= 
				\lim_{\theta \rightarrow 0+} 
				\frac{e^{-\frac{d}{\theta}}\left(d-ue^{\frac{d-u}{\theta}}\right)}
				{e^{-\frac{d}{\theta}}\left(1-e^{\frac{d-u}{\theta}}\right)} 
				= 
				d. \\
				\lim_{\theta \rightarrow \infty} 
				\mu_{\mbox{\tiny MTuM}}(\theta) 
				& = 
				\lim_{\theta \rightarrow \infty}\left[ \theta 
				+ \frac{de^{-\frac{d}{\theta}} 
					- ue^{-\frac{u}{\theta}}}{e^{-\frac{d}{\theta}} 
					- e^{-\frac{u}{\theta}}}\right] 
				\stackrel{y:=\frac{1}{\theta}} 
				=
				\lim_{y \rightarrow 0+} 
				\left[\frac{1}{y}+ \frac{de^{-dy}
					- ue^{-uy}}{e^{-dy} - e^{-uy}}\right] \\
				& = 
				\lim_{y \rightarrow 0+}\left[\frac{1}{y} 
				+ \frac{de^{(u-d)y}-u}{e^{(u-d)y}-1}\right] 
				= 
				\lim_{y \rightarrow 0+} 
				\left[\frac{e^{(u-d)y}-1+dye^{(u-d)y}-uy}
				{y\left( e^{(u-d)y}-1\right) }\right] \\
				& = 
				d - \lim_{y \rightarrow 0+}
				\frac{(u-d)-(u-d)e^{(u-d)y}}{e^{(u-d)y}-1+y(u-d)e^{(u-d)y}} \\ 
				& = 
				d - \lim_{y \rightarrow 0+}
				\frac{-(u-d)^{2}e^{(u-d)y}}{(u-d)e^{(u-d)y}+y(u-d)e^{(u-d)y}+(u-d)e^{(u-d)y}} \\
				& = 
				d + \frac{(u-d)^{2}}{u-d+u-d} 
				= 
				\frac{d+u}{2}. 
				\qedhere
			\end{align*} 
		\end{proof}
	\end{thm}
	
\begin{figure}[hbt!]
		\centering
		\includegraphics[width=1.00\textwidth]
		{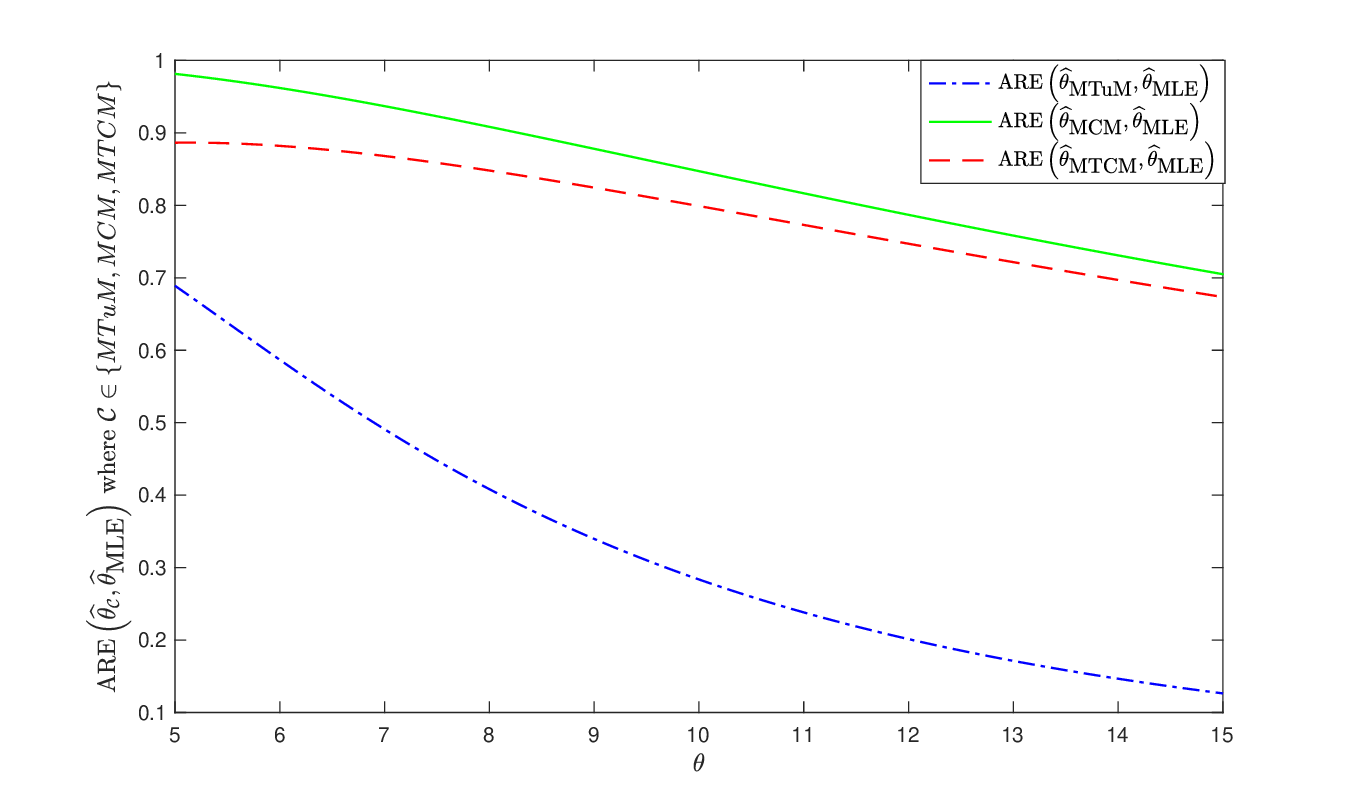}
		\caption{
			Graphs of 
			$
			\mbox{ARE}
			\left(
			\widehat{\theta}_{\mbox{\tiny $\mathcal{C}$}},
			\widehat{\theta}_{\mbox{\tiny MLE}}
			\right)
			$
			where 
			$
			\mathcal{C} 
			\in
			\{
			\mbox{MTuM, MCM, MTCM}
			\}
			$
			with 
			$
			(d,u) = (0.50, 23.00)
			$
			and 
			$\theta = 10$.
		}
		\label{fig:ARE_Graphs}
	\end{figure}

	Also from (\ref{eqn:pop_MTuM1}),
	we have
	\begin{align}
		\theta' 
		& :=  
		\frac{d\theta}{d\mu_{\mbox{\tiny MTuM}}}
		=
		\frac{p\theta^2}
		{de^{-\frac{d}{\theta}}(\theta+d)
			-ue^{-\frac{u}{\theta}}(\theta+u) 
			+ p\theta^2 
			- \mu_{\mbox{\tiny MTuM}} 
			\left(de^{-\frac{d}{\theta}} - ue^{-\frac{u}{\theta}}\right)} 
		\label{eqn:derOfThetaeMTuM1} \\
		& =
		\dfrac{p^{2}\theta^{2}}
		{p^{2}\theta^{2}-e^{-\frac{d+u}{\theta}}(u-d)^{2}}.
		\label{eqn:derOfThetaeMTuM2}
	\end{align}
	Therefore, by delta method, we get
	\begin{align}
		\label{eqn:MTuM_ThetaHat_AsymDist1}
		\widehat{\theta}_{\mbox{\tiny MTuM}} 
		\sim 
		\mathcal{AN}\left(\theta, \left(\theta'\right)^{2}
		\left( \frac{\sigma_{Y}^{2}}{np^{2}} 
		- \frac{(1-p)\mu_{Y}^{2}}{np^{3}}\right)\right)
		& = 
		\mathcal{AN}
		\left(\theta, 
		\dfrac{\theta^{2}}{n}
		\left(
		\dfrac{p\theta^{2}}
		{p^{2}\theta^{2}-e^{-\frac{d+u}{\theta}}(u-d)^{2}}
		\right)
		\right),
	\end{align}
	and hence, we have
	\begin{align}
		\label{eqn:MTuM_MLE_ARE_Exp1}
		\mbox{ARE}
		\left(
		\widehat{\theta}_{\mbox{\tiny MTuM}},
		\widehat{\theta}_{\mbox{\tiny MLE}}
		\right) 
		& = 
		\frac{\theta^{2}p^{3}}
		{\left(\theta'\right)^{2}
			\left(p\sigma_{Y}^{2}-(1-p)\mu_{Y}^{2}\right)}
		=
		\dfrac{p^{2}\theta^{2}-e^{-\frac{d+u}{\theta}}(u-d)^{2}}
		{p\theta^{2}}.
	\end{align}
	Clearly,
	$
	\mbox{ARE}
	\left(
	\widehat{\theta}_{\mbox{\tiny MTuM}},
	\widehat{\theta}_{\mbox{\tiny MLE}}
	\right) 
	$
	given by (\ref{eqn:MTuM_MLE_ARE_Exp1})
	is a function of the parameter $\theta$.
	Thus, it turns out that, 
	if we fix the left, $d$, and right, $u$, truncation thresholds
	and allow the tail probabilities, i.e., 
	$F(d \, | \, \theta)$ and $1-F(u \, | \, \theta)$,
	be random then the corresponding asymptotic relative
	efficiency is not stable 
	(see Figure \ref{fig:ARE_Graphs}).
	However, as in method of trimmed moments (MTM)
	\citep[see, e.g.,][]{MR2497558,zbg18a}
	if the tail probabilities 
	$F(d \, | \, \theta)$ and $1-F(u \, | \, \theta)$
	are fixed then we have the following result.
	
	\begin{prop}
		\label{prop:stabilityOfMTuM_ARE}
		Let $\theta_{1} \neq \theta_{2}$ be two exponential
		parameters with corresponding left and right truncation
		thresholds $d_{1}, \ d_{2}$ and $u_{1}, \ u_{2}$,
		respectively.
		Assume 
		$
		F(d_{1} \, | \, \theta_{1})
		= 
		F(d_{2} \, | \, \theta_{2})
		$
		and 
		$
		F(u_{1} \, | \, \theta_{1}) 
		= 
		F(u_{2} \, | \, \theta_{2}),
		$
		then it follows that 
		\begin{align}
			\label{equn:stabilityOfMTuM_ARE1}
			\mbox{ARE}
			\left(
			\widehat{\theta}_{\mbox{\tiny 1,MTuM}},
			\widehat{\theta}_{\mbox{\tiny 1,MLE}}
			\right) 
			& =
			\mbox{ARE}
			\left(
			\widehat{\theta}_{\mbox{\tiny 2,MTuM}},
			\widehat{\theta}_{\mbox{\tiny 2,MLE}}
			\right).
		\end{align}
		
		\begin{proof}
			A proof immediately follows from (\ref{eqn:MTuM_MLE_ARE_Exp1}).
		\end{proof}
	\end{prop}
	
	Numerical values of 
	$
	\mbox{ARE}
	\left(
	\widehat{\theta}_{\mbox{\tiny MTuM}},
	\widehat{\theta}_{\mbox{\tiny MLE}}
	\right) 
	$
	given by (\ref{eqn:MTuM_MLE_ARE_Exp1}) for 
	some selected values of left and right truncation
	thresholds $d$ and $u$, respectively are summarized
	on the first horizontal block of 
	Table \ref{table:CombinedAREs1}.
	
	As mentioned above, if $Y \sim$ 
	Pareto I $(\alpha, x_{0})$ with $x_{0}$ known then 
	$X:=\log{\left(\frac{Y}{x_{0}}\right)}
	\sim$ Exp$\left(\frac{1}{\alpha} =: \theta\right)$. 
	So, estimators of $\alpha$ of the single-parameter 
	Pareto distribution will share 
	the same AREs with estimators of $Exp(\theta)$, given that 
	$h(y) = \log{\left(\frac{y}{x_{0}}\right)}$. 
	The following result for single-parameter Pareto has been partially 
	derived in \cite{clark13}, but can easily be extended using 
	the tools of this section.
	
	\begin{thm} 
		\label{thm:gtTPI}
		Let $d$ and $u$ be the left and right truncation points, 
		respectively, for $Y \sim \text{Pareto I} \ (\alpha,x_{0})$. 
		Also, define 
		$A_{du} 
		:= 
		u^{\alpha}\left(1-\alpha \log{\left(\frac{x_{0}}{d}\right)}\right) 
		- d^{\alpha}\left(1-\alpha \log{\left(\frac{x_{0}}{u}\right)}\right)$ 
		and 
		$g_{du}(\alpha) 
		: = 
		\frac{A_{du}}{\alpha (u^{\alpha} 
			- d^{\alpha})}$. 
		Then the equation $\widehat{\mu}_{\mbox{\tiny MTuM}} 
		= 
		\mu_{\mbox{\tiny MTuM}}$ 
		has a unique solution provided that 
		$
		\displaystyle 
		\lim_{\alpha \rightarrow \infty} g_{du}(\alpha)
		< 
		\widehat{\mu}_{\mbox{\tiny MTuM}} 
		< 
		\lim_{\alpha \rightarrow 0+} g_{du}(\alpha).
		$
		\begin{proof}
			See Appendix \ref{apdx:proofs}.
		\end{proof} 
	\end{thm}
	
	Note that, given a truncated data, method of truncated moments 
	estimators for a normal population parameters can be 
	found in \cite{MR0045361} and \cite{MR0196848}.
	
	\subsection{Method of Fixed Censored Moments} 
	
	There are several versions of data censoring that occur in statistical modeling:
	interval censoring (it includes left and right censoring depending on which end
	point of the interval is infinite), type I censoring, type II censoring, and
	random censoring. For actuarial work, the most relevant type is {\em interval 
		censoring\/}. It occurs when complete sample observations are available within 
	some interval, say $(d, u]$, but data outside the interval is only partially 
	known. That is, counts are available but actual values are not. 
	That is, we observe the {\em i.i.d.\/} data
	\begin{equation}
		Z_{1}, \, Z_{2}, \ldots, Z_{n},
		\label{cdata}
	\end{equation}
	where each $Z$ is equal to the ground-up variable $X$, 
	if $X$ falls between 
	$d$ and $u$, and is equal to the corresponding end-point of
	the interval if 
	$X$ is beyond that point. 
	That is, $Z$ is given by
	\begin{align*}
		Z 
		& := 
		\min\big\{ \max (d, X), \, u \big\} 
		=
		d\ID\{X\leq d\} + X\ID\{d<X \leq u\} + u\ID\{X > u\}
		=
		\left\{ 
		\begin{array}{ll}
			d, & X \leq d; \\ 
			X, & d < X \leq u; \\
			u, & X > u.
		\end{array}
		\right.
	\end{align*}
	Therefore, instead of winsorizing fixed proportions
	of lowest and highest order statistics from an 
	observed sample \citep{zbg18a}, here we design 
	a method of fixed threshold censored moment for
	exponential distribution.
	
	Let 
	$X_{1},X_{2},\ldots,X_{n} \stackrel{i.i.d.}{\sim} \text{Exp}(\theta)$
	random variables. 
	Then, the sample censored mean is given by
	\[
	\widehat{\mu}_{\mbox{\tiny MCM}} 
	:= 
	\frac{d\sum_{i=1}^{n}\ID \{X_{i} \leq d\} 
		+ \sum_{i=1}^{n} X_{i}\ID \{d < X_{i} \leq u\} 
		+ u\sum_{i=1}^{n} \ID \{X_{i} > u\}}{n}.
	\]
	The corresponding population censored moments are:
	\begin{align*}
		\mu_{\mbox{\tiny MCM}} 
		& := 
		\E[Z] 
		= 
		d\left(1-e^{-\frac{d}{\theta}}\right)
		+\mu_{Y}+ue^{-\frac{u}{\theta}}, 
		\ \mbox{and} \\
		\mu_{\mbox{\tiny MCM,2}}
		& := 
		\E\left[Z^{2}\right] 
		= 
		d^2\left(1-e^{-\frac{d}{\theta}}\right)
		+\E\left[Y^2\right]+u^2e^{-\frac{u}{\theta}},
	\end{align*}
	where
	$Y := X\ID \{d < X \leq u\}$ 
	as in Section \ref{sec:MTuM_for_ExpPareto}.
	Thus, 
	$\sigma_{\mbox{\tiny MCM}}^{2} 
	= 
	\mu_{\mbox{\tiny MCM,2}}
	- 
	\mu_{\mbox{\tiny MCM}}^{2}$. 
	Moreover, setting 
	$\mu_{\mbox{\tiny MCM}} 
	=
	\widehat{\mu}_{\mbox{\tiny MCM}}$ 
	implies
	$d + \theta \left(e^{-\frac{d}{\theta}} 
	- e^{-\frac{u}{\theta}}\right) 
	= 
	\widehat{\mu}_{\mbox{\tiny MCM}},$
	which needs to be solved to get a 
	{\em method of censored moment} (MCM) 
	estimator, 
	$\widehat{\theta}_{\mbox{\tiny MCM}}$, 
	of $\theta$.
	
	\begin{thm}
		\label{thm:MCMExistenceTheorem}
		The equation 
		$\widehat{\mu}_{\mbox{\tiny MCM}} 
		=
		\mu_{\mbox{\tiny MCM}}$ 
		has a unique solution
		$\widehat{\theta}_{\mbox{\tiny MCM}}$ 
		provided that 
		$d < \widehat{\mu}_{\mbox{\tiny MCM}} < u$. 
		Otherwise, the solution does not exist.
		\begin{proof}
			A proof can similarly be established as in Theorem 
			\ref{thm:ExpTruncatedMean}.
		\end{proof}
	\end{thm}
	
	Moreover,  
	$
	\theta'
	:= 
	\frac{d\theta}{d\mu_{\mbox{\tiny MCM}}}
	=
	\frac{\theta}{p\theta 
		+ de^{-\frac{d}{\theta}} 
		- ue^{-\frac{u}{\theta}}}.
	$
	Then
	\begin{align}
		\label{eqn:MCM_ThetaHat_AsymDist1}
		\widehat{\theta}_{\mbox{\tiny MCM}}
		\sim
		\mathcal{AN}
		\left(
		\theta,
		\dfrac{1}{n}
		\dfrac{\theta^2 \sigma_{\mbox{\tiny MCM}}^{2}}
		{\left(p\theta+de^{-\frac{d}{\theta}}
			-ue^{-\frac{u}{\theta}}\right)^2}
		\right),
	\end{align}
	and hence
	\begin{align}
		\label{eqn:MCM_ARE1}
		\mbox{ARE}
		\left(\widehat{\theta}_{\mbox{\tiny MCM}},
		\widehat{\theta}_{\mbox{\tiny MLE}}
		\right) 
		& = 
		\frac{
			\left(p\theta+de^{-\frac{d}{\theta}}-ue^{-\frac{u}{\theta}}
			\right)^{2}}
		{\sigma_{\mbox{\tiny MCM}}^{2}}.
	\end{align}
	Again, as in MTuM case, see (\ref{eqn:MTuM_MLE_ARE_Exp1}),
	$
	\mbox{ARE}
	\left(\widehat{\theta}_{\mbox{\tiny MCM}},
	\widehat{\theta}_{\mbox{\tiny MLE}}
	\right) 
	$
	given by (\ref{eqn:MCM_ARE1})
	is a function of $\theta$
	and hence is not stable with respect to $\theta$.
	But if we fix the tail probabilities 
	$F(d \, | \, \theta)$ and 
	$1-F(u \, | \, \theta)$, 
	then we have the main result of this section.
	
	\begin{thm}
		\label{th:expCMTMEqual}
		For method of censored moments (MCM) with 
		$a = F(d|\theta)$ and 
		$b = 1-F(u|\theta)$, 
		then the following result holds:
		\[
		\mbox{ARE}
		\left(\widehat{\theta}_{\mbox{\tiny MCM}},
		\widehat{\theta}_{\mbox{\tiny MLE}}
		\right)
		=
		\mbox{ARE}
		\left(\widehat{\theta}_{\mbox{\tiny MTM}},
		\widehat{\theta}_{\mbox{\tiny MLE}}
		\right).
		\]
		
		\begin{proof}
			Following \cite{MR2497558}, we know that 
			\[
			\widehat{\theta\,}_{\mbox{\tiny MTM}} 
			\sim 
			\mathcal{AN}
			\left(\theta,\frac{\theta^2}{n}
			\Delta\right),\quad \text{with} \quad  \Delta 
			= 
			\frac{J(a,1-b)}{[I(a,1-b)]^2},
			\]
			where
			\[
			I(a,1-b)
			:=
			\int_{a}^{1-b}\log{(1-v)} \, dv 
			\quad \mbox{and} \quad 
			J(a,1-b) 
			:=  
			\int_{a}^{1-b} \int_{a}^{1-b} 
			\frac{\min\{v,w\}-vw}{(1-v)(1-w)} \, dw \,  dv.
			\]
			Therefore,
			\[
			\mbox{ARE}
			\left(
			\widehat{\theta}_{\mbox{\tiny MTM}},
			\widehat{\theta}_{\mbox{\tiny MLE}}
			\right)
			= 
			\frac{1}{\Delta} 
			= 
			\frac{[I(a,1-b)]^2}{J(a,1-b)}.
			\]
			On the other hand,
			\[
			\mbox{ARE}
			\left( 
			\widehat{\theta}_{\mbox{\tiny MCM}},
			\widehat{\theta}_{\mbox{\tiny MLE}}
			\right)
			=
			\frac{
				\left(
				p\theta+de^{-\frac{d}{\theta}}
				- ue^{-\frac{u}{\theta}}
				\right)^2
			}
			{\sigma_{\mbox{\tiny MCM}}^{2}}.
			\]
			So, we need to show that,
			\begin{align*}
				\mbox{ARE}
				\left( 
				\widehat{\theta}_{\mbox{\tiny MCM}},
				\widehat{\theta}_{\mbox{\tiny MLE}}
				\right)
				& = 
				\mbox{ARE}
				\left( 
				\widehat{\theta}_{\mbox{\tiny MTM}},
				\widehat{\theta}_{\mbox{\tiny MLE}}
				\right) 
				= 
				\frac{\theta^2 [I(a,1-b)]^2}{\sigma_{\mbox{\tiny MCM}}^{2}}.
			\end{align*}
			That is, 
			$J(a,1-b) 
			= 
			\dfrac{\sigma_{\mbox{\tiny MCM}}^{2}}{\theta^2}$. 
			For that, we have: 
			\begin{eqnarray*}
				\sigma_{\mbox{\tiny MCM}}^{2} 
				& = & 
				d^2
				\left(
				1-e^{-\frac{d}{\theta}}
				\right)
				+2\theta^2\left[\Gamma\left(3;\frac{d}{\theta}\right) - \Gamma\left(3;\frac{u}{\theta}\right)\right] 
				+ u^2e^{-\frac{u}{\theta}} -d^2 -2d\theta p-\theta^2 p^2 \\
				& = & 
				2\theta^2 (1-a-b) + 2\theta^2(-(1-a)\log{(1-a)} + b \log{(b)}) \\
				& & {} - \theta^{2}(1-a-b)(-2\log{(1-a)}+(1-a-b)). \\
				\frac{\sigma_{\mbox{\tiny MCM}}^{2}}{\theta^2} 
				& = &
				2(1-a-b)+2(b\log{(b)} - (1-a)\log{(1-a)}) \\
				& & {} - (1-a-b)(1-a-b-2\log{(1-a)}) \\
				& = & 
				(1-a-b)\left[a+\log{(1-a)}\right] - I(a,1-b) \\
				& & 
				{} + (1-b-1)\left[a-1+b+\log{\left(\frac{1-a}{b}\right)}\right] \\
				& = &
				(1-a-b)\left[a+\log{(1-a)}\right] - I(a,1-b) + (1-b-1)I_{1}(a,1-b) \\
				& = & 
				J(a,1-b).
			\end{eqnarray*}
			where 
			$\displaystyle I_{1}(a,1-b) 
			:= \int_{a}^{1-b}\frac{v}{1-v} \ dv
			= (a-1+b)+\log{\left(\frac{1-a}{b}\right)}$. 
		\end{proof}
	\end{thm}
	
	Here is an important and a new connection between 
	trimmed and interval censored population means for 
	any $F \in \mathscr{F}$, where $\mathscr{F}$ is
	the family of continuous parametric distributions.
	
	\begin{thm} 
		\label{thm:ifmtmmcm}
		Let 
		$F \in \mathscr{F}$ 
		be an arbitrary continuous ground-up 
		cumulative distribution function (cdf).
		Consider $d$ and $u$ be the lower and 
		upper thresholds, respectively. 
		Define 
		$a := F(d)$ and 
		$b := 1-F(u)$.
		Let 
		\[
		\mu_{\mbox{\tiny MCM}}
		=
		dF(d)
		+ \int_{d}^{u} z f(z) \, dz 
		+ u(1-F(u))
		\quad \mbox{and} \quad 
		\mu_{\mbox{\tiny MTM}}
		=
		\frac{1}{1-a-b}
		\int_{a}^{1-b}F^{-1}(v) \, dv,
		\]
		are, respectively, the fixed censored mean and 
		proportion trimmed mean of the same cdf $F$.
		Then 
		$IF(\mu_{\mbox{\tiny MCM}},x) 
		= 
		(1-a-b)IF(\mu_{\mbox{\tiny MTM}},x)$,
		$-\infty < x < \infty$
		where $IF$ stands for influence function.
		
		\begin{proof}
			The influence function of the trimmed mean is given by
			\citep[see, e.g.,][]{MR0362657, MR2488795}:
			\begin{align} \label{eq:ifmcm}
				IF(\mu_{\mbox{\tiny MTM}},x) 
				& =
				\frac{1}{1-a-b}  \int_{a}^{1-b}
				\left.\left(\frac{d}{d\lambda}
				F_{\lambda}^{-1}(v)\right)\right|_{\lambda = 0} \, dv 
				=
				\frac{1}{1-a-b} 
				\int_{a}^{1-b}\frac{v 
					- \ID\{F(x) \leq v\}}{f(F^{-1}(v))} \, dv,
			\end{align} 
			where 
			$F_{\lambda}
			:= 
			(1-\lambda)F + \lambda \delta_{x}$
			and $\delta_{x}$ is the point mass at $x$.
			Since $d$ and $u$ are left and right censored points, respectively. 
			Then, the censored mean is:
			\begin{align*}
				\mu_{\mbox{\tiny MCM}}[F]
				= 
				\int_{d}^{u}z \, dF_{Z}(z) 
				& = 
				dF(d) + \E[X \ID\{d < X < u\}] + u(1-F(u)) \\
				& =
				dF(d) + \int_{d}^{u}z \, dF(z) + u(1-F(u)),
			\end{align*}
			where $F_{Z}$ is the  distribution function
			of $Z$ given by:
			\begin{equation}
				F_{Z}(z \, | \, d, u)
				= 
				\mathbb{P}
				\left[ \min\big\{ \max (d, X), \, u \big\} \leq z \right] 
				= 
				\left\{ 
				\begin{array}{ll}
					0, & z < d; \\
					F(z), & d \leq z < u; \\
					1,  & z \geq u, \\
				\end{array}
				\right.
				\label{ccdf}
			\end{equation}
			Further, 
			$
			\mu_{\mbox{\tiny MCM}}[F_{\lambda}] 
			= 
			dF_{\lambda}(d) + \int_{d}^{u}z \, dF_{\lambda}(z) 
			+ u(1-F_{\lambda}(u)).
			$
			Note that the influence function is just a special case of first 
			order G{\^a}teaux derivative \cite[see, e.g.,][Section 2.3]{MR829458}. 
			Thus, a simpler computational formula to get the IF is
			\citep[see, e.g.,][Chapter 6]{MR595165}:
			\begin{equation*}
				\label{eq:IFcomp}
				IF(\mu_{\mbox{\tiny MCM}},x) 
				= 
				\left.
				\frac{d\mu_{\mbox{\tiny MCM}}[F_{\lambda}]}{d\lambda}\right|_{\lambda=0}.
			\end{equation*}
			It is clear that 
			$\left.\frac{dF_{\lambda}(d)}{d\lambda}\right|_{\lambda=0} 
			=
			-F(d)+\delta_{x}(d)$ 
			and similarly 
			$\left.\frac{d(1-F_{\lambda}(u))}{d\lambda}\right|_{\lambda=0} 
			=
			F(u)-\delta_{x}(u)$.
			Also, by using Leibniz's rule for differentiation under integral sign, we get 
			\begin{eqnarray*}
				\frac{d}{d\lambda}\int_{d}^{u}z \, dF_{\lambda}(z) 
				& = & 
				\frac{d}{d\lambda}
				\int_{F_{\lambda}(d)}^{F_{\lambda}(u)}F_{\lambda}^{-1}(v) \, dv \\
				& = & 
				F_{\lambda}^{-1}(F_{\lambda}(u)) \frac{d}{d\lambda}F_{\lambda}(u) - F_{\lambda}^{-1}(F_{\lambda}(d)) \frac{d}{d\lambda}F_{\lambda}(d) + \int_{F_{\lambda}(d)}^{F_{\lambda}(u)}\frac{d}{d\lambda}F_{\lambda}^{-1}(v) \, dv \\
				& = & 
				u \frac{d}{d\lambda}F_{\lambda}(u) - d \frac{d}{d\lambda}F_{\lambda}(d) + \int_{F_{\lambda}(d)}^{F_{\lambda}(u)}\frac{d}{d\lambda}F_{\lambda}^{-1}(v) \, dv. \\
				\left. 
				\frac{d}{d\lambda}\int_{d}^{u}z \, dF_{\lambda}(z)\right|_{\lambda = 0} 
				& = &
				u\left.\frac{dF_{\lambda}(u)}{d\lambda}\right|_{\lambda=0} - d\left.\frac{dF_{\lambda}(d)}{d\lambda}\right|_{\lambda=0} + \left.\int_{F_{\lambda}(d)}^{F_{\lambda}(u)}\frac{d}{d\lambda}
				F_{\lambda}^{-1}(v)\, dv \right|_{\lambda = 0} \\
				& = & 
				u\left.\frac{dF_{\lambda}(u)}{d\lambda}\right|_{\lambda=0} - d\left.\frac{dF_{\lambda}(d)}{d\lambda}\right|_{\lambda=0} 
				+ \int_{a}^{1-b}\left.\left(\frac{d}{d\lambda}
				F_{\lambda}^{-1}(v)\right)\right|_{\lambda = 0} \, dv.
			\end{eqnarray*}
			Therefore,
			\begin{eqnarray} \label{eq:ifmtm}
				IF(\mu_{\mbox{\tiny MCM}},x) 
				& = & 
				\left.\frac{d\mu_{\mbox{\tiny MCM}}
					[F_{\lambda}]}{d\lambda}\right|_{\lambda=0} \nonumber \\
				& = & 
				d\left.\frac{dF_{\lambda}(d)}{d\lambda}\right|_{\lambda=0} + u\left.\frac{d(1-F_{\lambda}(u))}{d\lambda}\right|_{\lambda=0} 
				+ \left. \frac{d}{d\lambda}\int_{d}^{u}y \, 
				dF_{\lambda}(y)\right|_{\lambda = 0} \nonumber \\
				& = & 
				d\left.\frac{dF_{\lambda}(d)}{d\lambda}\right|_{\lambda=0} + u\left.\frac{d(1-F_{\lambda}(u))}{d\lambda}\right|_{\lambda=0} + u\left.\frac{dF_{\lambda}(u)}{d\lambda}\right|_{\lambda=0} \nonumber \\
				& & {}  
				- d\left.\frac{dF_{\lambda}(d)}{d\lambda}\right|_{\lambda=0} + \int_{a}^{1-b}\left.\left(\frac{d}{d\lambda}
				F_{\lambda}^{-1}(v)\right)\right|_{\lambda = 0} \ dv \nonumber \\
				& = & 
				\int_{a}^{1-b}\left.\left(\frac{d}{d\lambda}
				F_{\lambda}^{-1}(v)\right)\right|_{\lambda = 0} \ dv.
			\end{eqnarray}
			Thus, from Equations (\ref{eq:ifmtm}) and (\ref{eq:ifmcm}), 
			$IF(\mu_{\mbox{\tiny MCM}},x) 
			= 
			(1-a-b)IF(\mu_{\mbox{\tiny MTM}},x)$.
		\end{proof}
	\end{thm}
	
	\begin{figure}[hbt!]
		\flushright{\includegraphics[width=16.5cm,height=9.0cm,keepaspectratio]
			{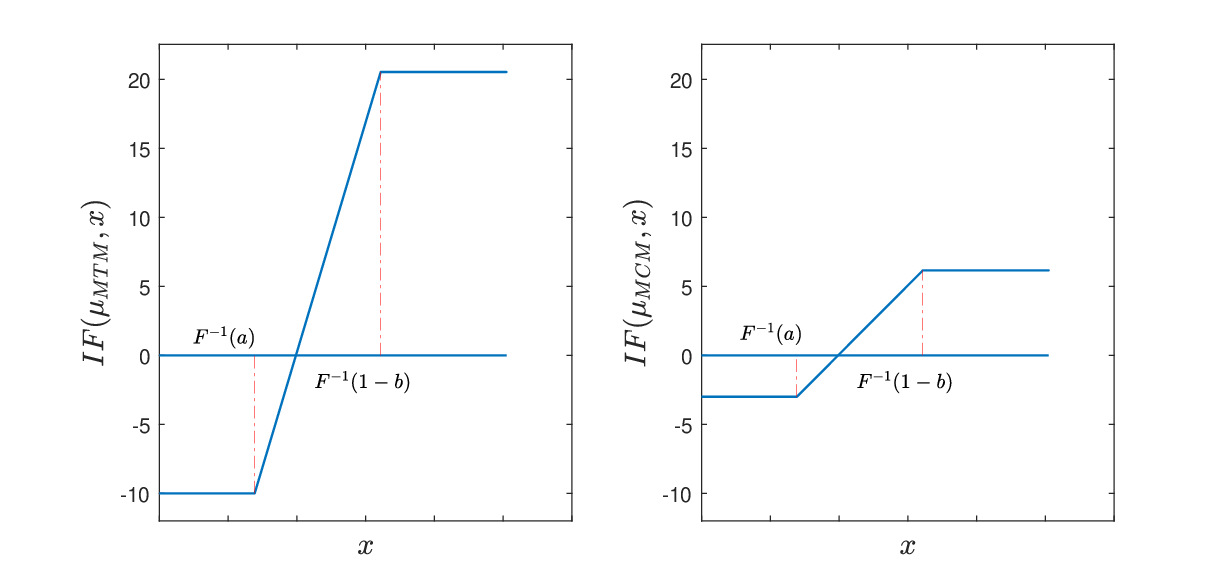}}
		\caption{Influence functions of trimmed mean (left panel) 
			and censored mean (right panel).}
		\label{fig:MTM_MCM_IFs}
	\end{figure}
	
	\noindent 
	The following two points are immediate consequence of
	Theorem \ref{thm:ifmtmmcm}.
	\begin{enumerate}[label=(\roman*)]
		\item 
		Censored mean is asymptotically more stable than trimmed mean. 
		This point is more clear from the Figure \ref{fig:MTM_MCM_IFs} as 
		the graph of $IF(\mu_{\mbox{\tiny MCM}},x)$ is just the vertical 
		contraction of the graph of $IF(\mu_{\mbox{\tiny MTM}},x)$ by the 
		contracting factor $0 < 1-a-b < 1$ with the assumption $0 < a+b < 1$. 
		\item 
		The asymptotic investigation of censored mean could be quite 
		challenging due to thresholds. 
		So, in this situation one can assess 
		the asymptotic distributional properties of censored mean 
		through the corresponding properties of trimmed mean as the
		asymptotic variance of an estimator is the expectation 
		of the square of the corresponding IF 
		\citep[see, e.g.,][]{MR0362657, MR3839299}. 
	\end{enumerate}
	
	\subsection{Method of Actuarial Payment--type Moments}
	\label{sec:loss_data_MTCM}
	
	Insurance contracts have coverage modifications that need to be taken into 
	account when modeling the underlying loss variable. Usually the coverage
	modifications such as deductibles, policy limits, and coinsurance are 
	introduced as loss control strategies so that unfavorable policyholder 
	behavioral effects (e.g., adverse selection) can be minimized. 
	Therefore, the actuarial loss data are left truncated and right censored
	in nature. Motivated with this nature of the loss data, 
	here we design an estimation approach, called 
	{\em insurance payment--type estimators} and is basically left truncated
	and right censored method of moments.
	
	Let $X_{1},\ldots,X_{n}$ be {\em i.i.d.\/} 
	random variables with common exponential cdf $F(\cdot|\theta)$.  
	Define the left- truncated (at $d$) and right-censored (at $u$) 
	sample moment as:
	\begin{align*}
		\widehat{\mu}_{\mbox{\tiny MTCM}} 
		& := 
		\frac{\sum_{i=1}^{n} X_{i}
			\ID \{d < X_{i} \leq u\} + u\sum_{i=1}^{n} 
			\ID \{X_{i} > u\}}{\sum_{i=1}^{n} \ID \{X_{i} > d\}} 
		= 
		\frac{\overline{W}_{n}}{\tau_{n}}.
	\end{align*}
	where 
	$
	W 
	:= 
	X\ID\{d<X \leq u\}+u\ID\{X > u\}$,
	$\tau_{n} = 1-F_{n}(d)$, and $\tau = 1-F(d|\theta)$. 
	The covariance of $W$ and $\tau_{1}$ is given as
	$
	\sigma_{W\tau_{1}}^{2} 
	= 
	\mathbb{C}{ov}\left(W,\tau_{1}\right) 
	= 
	\mu_{W}(1-\tau),
	$
	with 
	\begin{align*}
		\mu_{W}
		=
		\E\left[W\right]  
		= 
		\mu_{Y} +u(1-F(u|\theta)) 
		\quad \text{and} \quad
		\E\left[{W}^{2}\right] 
		=
		\E\left[Y^{2}\right] 
		+ u^{2}(1-F(u|\theta)),
	\end{align*}
	where
	$Y := X\ID \{d < X \leq u\}$ 
	as in Section \ref{sec:MTuM_for_ExpPareto}.
	Then, by multivariate Central Limit Theorem, we have
	\[
	\left(\overline{W}_{n},\tau_{n}\right) 
	\sim \mathcal{AN}
	\left((\mu_{W},\tau),
	\frac{1}{n}
	\begin{bmatrix}
	\sigma_{W}^{2} & \sigma_{W\tau_{1}}^{2} \\
	\sigma_{W\tau_{1}}^{2} & \tau(1-\tau)
	\end{bmatrix}
	\right).
	\]
	Then, by delta method with a function 
	$g(x_{1},x_{2}) := \frac{x_{1}}{x_{2}}$, 
	$x_{2} \neq 0$,
	we have
	\[
	\widehat{\mu}_{\mbox{\tiny MTCM}} 
	=
	\frac{\overline{W}_{n}}{\tau_{n}} 
	\sim 
	\mathcal{AN}\left(\frac{\mu_{W}}{\tau},
	\frac{1}{n}
	\left(\frac{\sigma_{W}^{2}}{\tau^2}
	- \frac{\mu_{W}^{2}(1-\tau)}{\tau^3} 
	\right)\right).
	\]
	The population version of
	$\widehat{\mu}_{\mbox{\tiny MTCM}}$ is given by
	\begin{align*}
		{} & \mu_{\mbox{\tiny MTCM}} 
		= 
		\frac{\E[W]}{1-F(d|\theta)}
		=
		\frac{\theta\left(e^{-\frac{d}{\theta}} 
			- e^{-\frac{u}{\theta}} \right)
			+de^{-\frac{d}{\theta}}}{e^{-\frac{d}{\theta}}} 
		=
		\dfrac{p \theta + d \tau}{\tau} \\
		\Rightarrow \quad & 
		\theta' 
		:=
		\frac{d\theta}{d\mu_{\mbox{\tiny MTCM}}}
		=
		\frac{\tau\theta^2}{p \theta^{2}
			+ \theta\left(de^{-\frac{d}{\theta}}
			- ue^{-\frac{u}{\theta}} \right) + d^{2}\tau
			- d\tau\mu_{\mbox{\tiny MTCM}}} 
		=
		\dfrac{\tau \theta^{2}}
		{p \theta^{2} - e^{-\frac{u}{\theta}}(u-d)}.
	\end{align*}
	
	A solution, if exists, of the equation $\widehat{\mu}_{\mbox{\tiny MTCM}} 
	=
	\mu_{\mbox{\tiny MTCM}}$, say 
	$\widehat{\theta}_{\mbox{\tiny MTCM}}$,
	is called the 
	{\em method of truncated and censored moment} (MTCM)
	estimator of $\theta$. 
	Let $b := e^{-\frac{u}{\theta}}$,
	then by delta method the asymptotic 
	distribution 
	and ARE are, respectively, given by
	\begin{align}
		\widehat{\theta}_{\mbox{\tiny MTCM}}
		&\sim
		\mathcal{AN}\left(\theta,
		\frac{(\theta')^{2}}{n}
		\left(\frac{\sigma_{W}^{2}}{\tau^2}
		- \frac{(1-\tau)\mu_{W}^{2}}{\tau^3} 
		\right)\right)
		=
		\mathcal{AN}
		\left(\theta,
		\frac{\theta^{2}}{n}
		\left(
		\dfrac{p\theta^{2}
			\left(\tau+b\right)
			-2\theta \tau b(u-d)}
		{\tau 
			\left[ 
			p\theta-b
			\left( \frac{u-d}{\theta} \right)
			\right]^{2}}
		\right)
		\right) 
		\label{eqn:LTRC_AsyDist1} 
	\end{align}
	\begin{align}
		\mbox{ARE}
		\left(
		\widehat{\theta}_{\mbox{\tiny MTCM}},
		\widehat{\theta}_{\mbox{\tiny MLE}}
		\right) 
		& = 
		\frac{\theta^{2}\tau^{3}}{\left(\theta'\right)^{2}
			\left(\tau\sigma_{W}^{2}-(1-\tau)
			\mu_{W}^{2}\right)}
		= 
		\dfrac{
			\left[ 
			p-b\left( \frac{u-d}{\theta}\right)
			\right]^{2}}
		{p(1+b/\tau)
			-2b\left( \frac{u-d}{\theta}\right)
		}.
		\label{eqn:LTRC_ARE1}
	\end{align}
	Similar to 
	(\ref{eqn:MTuM_MLE_ARE_Exp1}) and 
	(\ref{eqn:MCM_ARE1}), 
	$
	\mbox{ARE}
	\left(
	\widehat{\theta}_{\mbox{\tiny MTCM}},
	\widehat{\theta}_{\mbox{\tiny MLE}}
	\right) 
	$
	given by (\ref{eqn:LTRC_ARE1})
	is a function of $\theta$.
	But if we fix the tail probabilities then 
	we have the following stability result.
	
	\begin{prop}
		\label{prop:stabilityOfLTRC_ARE}
		Let $\theta_{1} \neq \theta_{2}$ be two exponential
		parameters with corresponding left and right truncation
		thresholds $d_{1}, \ d_{2}$ and $u_{1}, \ u_{2}$,
		respectively.
		Assume 
		$
		F(d_{1} \, | \, \theta_{1})
		= 
		F(d_{2} \, | \, \theta_{2})
		$
		and 
		$
		F(u_{1} \, | \, \theta_{1}) 
		= 
		F(u_{2} \, | \, \theta_{2}),
		$
		then it follows that 
		\begin{align}
			\label{equn:stabilityOfMTCM_ARE1}
			\mbox{ARE}
			\left(
			\widehat{\theta}_{\mbox{\tiny 1,MTCM}},
			\widehat{\theta}_{\mbox{\tiny 1,MLE}}
			\right) 
			& =
			\mbox{ARE}
			\left(
			\widehat{\theta}_{\mbox{\tiny 2,MTCM}},
			\widehat{\theta}_{\mbox{\tiny 2,MLE}}
			\right).
		\end{align}
		
		\begin{proof}
			With the assumptions given, we have 
			\[
			e^{-\frac{d_{1}}{\theta_{1}}}
			=
			e^{-\frac{d_{2}}{\theta_{2}}}, \
			e^{-\frac{u_{1}}{\theta_{1}}}
			=
			e^{-\frac{u_{2}}{\theta_{2}}}
			\quad \implies \quad 
			\dfrac{u_{1}-d_{1}}{\theta_{1}} 
			=
			\dfrac{u_{2}-d_{2}}{\theta_{2}},
			\]
			and then the conclusion follows directly from 
			(\ref{eqn:LTRC_ARE1}).
		\end{proof}
	\end{prop}
	
	
	\begin{table}[hbt!] 
		\centering
		\caption{
			Numerical values of 
			$
			\mbox{ARE}
			\left(
			\widehat{\theta}_{\mbox{\tiny $\mathcal{C}$}},
			\widehat{\theta}_{\mbox{\tiny MLE}}
			\right)$ 
			where
			$
			\mathcal{C} 
			\in 
			\{
			\mbox{MTuM, MCM, MTCM} 
			\}
			$, 
			respectively, given by 
			(\ref{eqn:MTuM_MLE_ARE_Exp1}), 
			(\ref{eqn:MCM_ARE1}), and 
			(\ref{eqn:LTRC_ARE1})
			for various values of left and right truncation
			thresholds $d$ and $u$ from $Exp(\theta = 10)$.
			The truncation  thresholds $d$ and $u$ are rounded
			to two decimal places; for example, 
			$0.51 \approx F^{-1}(0.05), 
			\ 18.97 \approx F^{-1}(0.85)$, etc.}
		\label{table:CombinedAREs1}
		\begin{tabular}{cc|cccccccc}
			\hline \\[-3.00ex]
			\multicolumn{2}{c|}{} &		
			\multicolumn{8}{c}{u{\tiny $(1-F(u|\theta))$}} \\
			{} & d{\tiny $(F(d|\theta))$} & {\scriptsize~$\infty${\tiny (.00)}} &
			{\scriptsize~29.96{\tiny (.05)}} & {\scriptsize~23.03{\tiny (.10)}} & 
			{\scriptsize~18.97{\tiny (.15)}} & {\scriptsize~13.86{\tiny (.25)}} &
			{\scriptsize~7.13{\tiny (.49)}} & {\scriptsize~3.57{\tiny (.70)}} & 
			{\scriptsize~1.63 {\tiny (.85)}} \\
			\hline\hline
			& & & & & & & & & \\[-2.25ex]
			\multirow{8}{*}{\rotatebox{90}{$\mbox{ARE}
					\left(
					\widehat{\theta}_{\mbox{\tiny MTuM}},
					\widehat{\theta}_{\mbox{\tiny MLE}}
					\right)$}} &
			{\scriptsize~0{\tiny (.00)}} & 
			1 & .478 & .311 & .215 & .109 &	.021 & .003 & .000 \\
			& {\scriptsize~0.51{\tiny (.05)}} &
			.950 & \fbox{.443} & .284 & .193 & .095 & .016 & .002 & .000 \\
			& {\scriptsize~1.05{\tiny (.10)}} & 
			.900 & .408 & .257 & .172 & .082 & .012 & .001 & .000 \\
			& {\scriptsize~1.63{\tiny (.15)}} & 
			.850 & .373 & .231 & .152 & .069 & .009 & .000 & - \\
			& {\scriptsize~2.88{\tiny (.25)}} & 
			.750 & .307 & .182 & .114 & .047 & .004 & .000 & - \\
			& {\scriptsize~6.73{\tiny (.49)}} & 
			.510 & .161 & .080 & .042 & .011 & .000 & - & - \\
			& {\scriptsize~12.04{\tiny (.70)}} & 
			.300 & .057 & .019 & .006 & .000 & - & - & - \\
			& {\scriptsize~18.97{\tiny (.85)}} & 
			.150 & .009 & .001 & - & - & - & - & - \\
			\hline\hline
			& & & & & & & & & \\[-2.25ex]
			\multirow{8}{*}{\rotatebox{90}{
					$
					\mbox{ARE}
					\left(
					\widehat{\theta}_{\mbox{\tiny MCM}},
					\widehat{\theta}_{\mbox{\tiny MLE}}
					\right)$}} & 
			{\scriptsize~0{\tiny (.00)}} & 
			1 & .918 & .847 & .783 & .666 &	.423 & .238 & .116 \\
			& {\scriptsize~0.51{\tiny (.05)}} &
			1 & \fbox{.918} & .848 & .783 & .667 & .425 & .242 & .122 \\
			& {\scriptsize~1.05{\tiny (.10)}} & 
			1 & .918 & .848 & .785 & .669 & .430 & .250 & .135 \\
			& {\scriptsize~1.63{\tiny (.15)}} & 
			.999 & .918 & .850 & .787 & .672 & .436 & .261 & - \\
			& {\scriptsize~2.88{\tiny (.25)}} & 
			.995 & .918 & .851 & .790 & .679 & .452 & .285 & - \\
			& {\scriptsize~6.73{\tiny (.49)}} & 
			.958 & .897 & .839 & .786 & .688 & .487 & - & - \\
			& {\scriptsize~12.04{\tiny (.70)}} & 
			.857 & .824 & .781 & .738 & .659 & - & - & - \\
			& {\scriptsize~18.97{\tiny (.85)}}& 
			.681 & .688 & .663 & - & - & - & - & - \\
			\hline\hline
			& & & & & & & & & \\[-2.25ex]
			\multirow{8}{*}{\rotatebox{90}{$\mbox{ARE}
					\left(
					\widehat{\theta}_{\mbox{\tiny MTCM}},
					\widehat{\theta}_{\mbox{\tiny MLE}}
					\right)$}} & 
			{\scriptsize~0{\tiny (.00)}} & 
			1 & .918 & .847 & .783 & .666 &	.423 & .238 & .116 \\
			& {\scriptsize~0.51{\tiny (.05)}} &
			.950 & \fbox{.868} & .798 & .735 & .619 & .380 & .197 & .077 \\
			& {\scriptsize~1.05{\tiny (.10)}} & 
			.900 & .819 & .750 & .687 & .572 & .336 & .157 & .038 \\
			& {\scriptsize~1.63{\tiny (.15)}} & 
			.850 & .768 & .700 & .638 & .525 & .292 & .116 & - \\
			& {\scriptsize~2.88{\tiny (.25)}} & 
			.750 & .670 & .603 & .542 & .432 & .208 & .038 & - \\
			& {\scriptsize~6.73{\tiny (.49)}} & 
			.510 & .434 & .371 & .315 & .216 & .015 & - & - \\
			& {\scriptsize~12.04{\tiny (.70)}} & 
			.300 & .229 & .173 & .124 & .039 & - & - & - \\
			& {\scriptsize~18.97{\tiny (.85)}}& 
			.150 & .087 & .040 & - & - & - & - & - \\
			\hline\hline
		\end{tabular}
	\end{table}
	
	From Table \ref{table:CombinedAREs1}, 
	it follows evidently that 
	\[
	\mbox{ARE}
	\left(
	\widehat{\theta}_{\mbox{\tiny MTuM}},
	\widehat{\theta}_{\mbox{\tiny MLE}}
	\right)
	\le 
	\mbox{ARE}
	\left(
	\widehat{\theta}_{\mbox{\tiny MTCM}},
	\widehat{\theta}_{\mbox{\tiny MLE}}
	\right)
	\le
	\mbox{ARE}
	\left(
	\widehat{\theta}_{\mbox{\tiny MCM}},
	\widehat{\theta}_{\mbox{\tiny MLE}}
	\right).
	\]
	This inequality is intuitive because 
	MTuM is more robust than MCM and MTCM.
	As a result, MTuM estimators lose more efficiency
	and converge to the asymptotic results slower.
	For example, if the lower and upper truncation 
	thresholds are, respectively, 
	$d = 0.51$ and  $u = 29.96$ then 
	$
	\mbox{ARE}
	\left(
	\widehat{\theta}_{\mbox{\tiny MTuM}},
	\widehat{\theta}_{\mbox{\tiny MLE}}
	\right) 
	=
	0.443
	$
	and
	$
	\mbox{ARE}
	\left(
	\widehat{\theta}_{\mbox{\tiny MCM}},
	\widehat{\theta}_{\mbox{\tiny MLE}}
	\right)
	=
	0.918
	$. 
	That is, we lose approximately 52\% 
	efficiency by going from MCM to MTuM. 
	The reason that MTuM relative efficiency is much lower 
	than the corresponding 
	MCM is that the censored sample size is always fixed
	but even if we fix the  truncation thresholds, 
	the truncated sample size is random.
	Further, MTuM disregards the observations beyond the 
	truncation thresholds in order to control the influence 
	of extremes in statistical inference. 
	MCM controls such influence of extremes differently, 
	i.e., those observations which  are beyond the thresholds 
	are adjusted to be equal to the corresponding thresholds 
	and hence increases the efficiency significantly.
	MTCM controls the influence of extremes by disregarding 
	the observations below lower threshold and adjusting 
	the observations above upper threshold to be equal to the 
	upper threshold which makes the MTCM entries in between 
	the corresponding MTuM and MCM entries.
	Due to Theorem \ref{th:expCMTMEqual},
	entries for 
	$
	\mbox{ARE}
	\left(
	\widehat{\theta}_{\mbox{\tiny MCM}},
	\widehat{\theta}_{\mbox{\tiny MLE}}
	\right)
	$
	are identical to 
	$
	\mbox{ARE}
	\left(
	\widehat{\theta}_{\mbox{\tiny MTM}},
	\widehat{\theta}_{\mbox{\tiny MLE}}
	\right)
	$ 
	entries found in \citep[][Table 1]{MR2497558}.
	
	\begin{thm}
		\label{thm:MTCMExistenceTheorem}
		The equation $\widehat{\mu}_{\mbox{\tiny MTCM}} 
		=
		\mu_{\mbox{\tiny MTCM}}$ 
		has a unique solution $\widehat{\theta}_{\mbox{\tiny MTCM}}$
		provided that $d < \widehat{\mu}_{\mbox{\tiny MTCM}} < u$.
		Otherwise, the solution does not exist.
		\begin{proof}
			A proof can similarly be established as in Theorem 
			\ref{thm:ExpTruncatedMean}.
		\end{proof}
	\end{thm}
	
	\section{Simulation Study}
	\label{sec:SimStudy}
	
	This section supplements the theoretical results 
	we developed in Section \ref{sec:MTuM_ExpPareto} via simulation. 
	The main goal is to access the size of the sample such that the
	estimators are free from bias 
	(given that the estimators are asymptotically unbiased),
	justify the asymptotic normality, and
	their finite sample relative efficiencies (RE) are 
	approaching to the corresponding AREs. 
	To compute RE of different estimators (MTuM, MCM, and MTCM) 
	we use MLE as a benchmark. 
	Thus, the definition of asymptotic relative 
	efficiency given by equation 
	(\ref{eq:infinite_relative_efficiency_benchmark_MLE})
	for finite sample performance translates to:
	\begin{equation*} 
		\label{eq:finite_relative_efficiency_benchmark_MLE}
		RE(\mathcal{C}, MLE) 
		=
		\frac{\text{asymptotic variance of MLE estimator}}
		{\text{small-sample variance of a competing estimator }
			\mathcal{C}},
	\end{equation*}
	where the denominator is the empirical 
	mean square error matrix of the 
	competing estimator $\mathcal{C}$. 
	
	\begin{figure}[hbt!]
		\centering
		\includegraphics[width=1.00\textwidth]
		{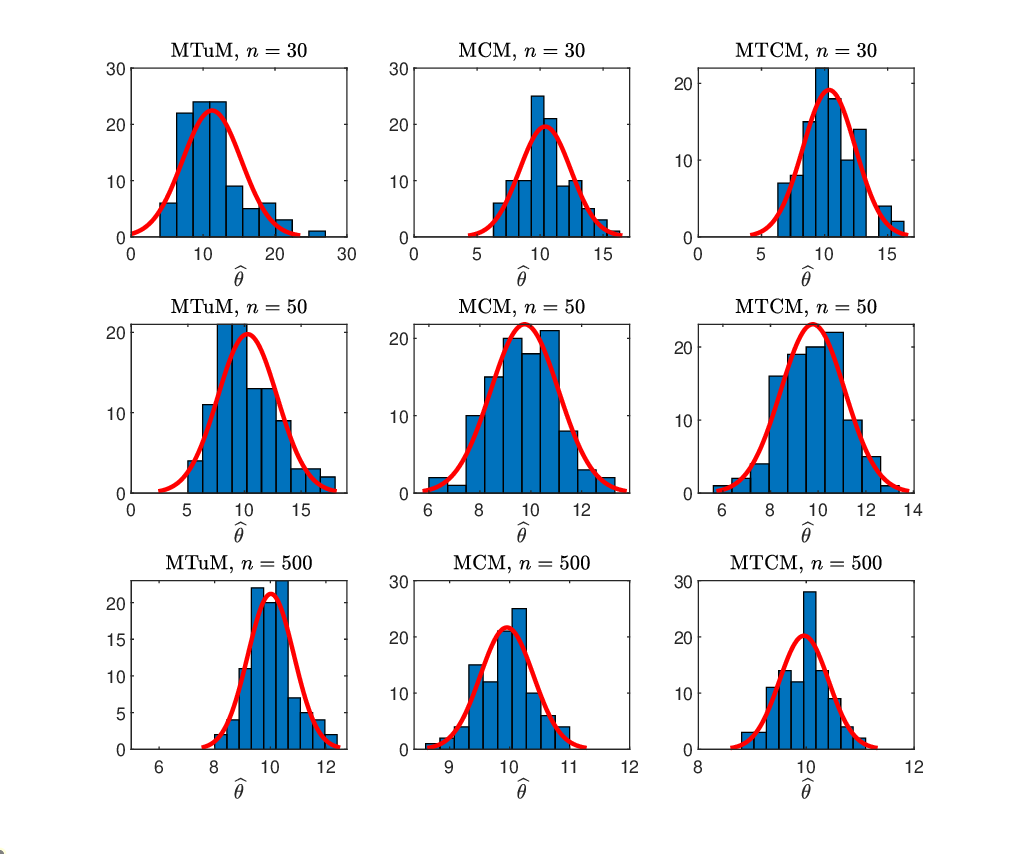}
		\caption{Histograms of 100 estimated values of the 
			parameter $\mbox{Exp}(\theta = 10)$ via MTuM, MCM, 
			and MTCM with
			$(d,u) = (0.50,23.00)$ 
			and sample sizes 
			$n = 30, \ 50, \ 500$.}
		\label{fig:asymDistGraph}
	\end{figure}
	
	From $\mbox{Exp}(\theta = 10)$,
	we first monitor the approximate normality 
	distributional properties of the MTuM, MCM, and
	MTCM estimators of $\theta$ 
	given, respectively, by 
	(\ref{eqn:MTuM_ThetaHat_AsymDist1}), 
	(\ref{eqn:MCM_ThetaHat_AsymDist1}), and 
	(\ref{eqn:LTRC_AsyDist1})
	with 
	$(d,u) = (0.50,23.00)$ and finite
	sample sizes $n = 30, \ 50$.
	We generate 100 samples for each sample size
	$n = 30, \ 50$ and estimate the values of 
	$\theta$ from each sample via MTuM, MCM, and MTCM. 
	We plot the histograms of those 100 estimated values 
	of $\theta$ in Figure \ref{fig:asymDistGraph}.
	Clearly, the histograms corresponding to MTuM 
	(for $n = 30, \ 50$)
	are positively skewed
	(but turns out to be symmetric for $n = 500$)
	and hence the asymptotic
	normality property of 
	$
	\widehat{\theta}_{\mbox{\tiny MTuM}}
	$
	given by 
	(\ref{eqn:MTuM_ThetaHat_AsymDist1})
	can be achieved slower, i.e.,
	only for bigger sample sizes, than MCM and MTCM.
	On the other hand the asymptotic normality property of 
	$
	\widehat{\theta}_{\mbox{\tiny MCM}}
	$
	given by 
	(\ref{eqn:MCM_ThetaHat_AsymDist1}) and
	$
	\widehat{\theta}_{\mbox{\tiny MTCM}}
	$
	given by
	(\ref{eqn:LTRC_AsyDist1})
	can be justified even for the sample of size 
	$n = 30$. 
	
	Second, again from exponential distribution 
	$F(\cdot|\theta=10)$ we generate $10,000$ 
	samples of a specified length $n$ using Monte Carlo. 
	For each sample we estimate the parameter of $F$ 
	using various MTuM, MCM, 
	and MTCM estimators and then compute the average mean and 
	RE of those $10,000$ estimates. 
	This process is repeated 
	$10$ times and the $10$ average means and the $10$ RE's 
	are again averaged and their standard errors are reported. 
	Such repetitions are useful for assessing standard errors 
	of the estimated means and RE's. 
	Hence, our findings are essentially based on $100,000$ samples.
	The standardized ratio
	$\widehat{\theta}/\theta$
	\label{mis:Mean1}
	that we report is defined as the 
	average of $100,000$ estimates divided by the true value
	of the parameter that we are estimating. 
	We observe the performance of different methods of 
	estimation for exponential distribution 
	(see Section \ref{sec:MTuM_ExpPareto}) in the aspects of
	\begin{enumerate}[label=(\roman*)]
		\setlength{\itemsep}{-2mm}
		\item 
		Sample size: $n = 50, \ 100, \ 250, \ 500, \ 1000$.
		\item 
		\label{item:EstimatorsOfTheta}
		Estimators of $\theta$:
		\begin{enumerate}[label=(\alph*)]
			\item 
			MLE, which is a special case of all others. 
			\item 
			MTuM, MCM, MTCM.
			\item 
			For the selected proportions
			$a = b = 0$; 
			$a = b = 0.05$; 
			$a = b = 0.10$;
			$a = b = 0.15$; 
			$a = b = 0.25$; 
			$a = 0.10$ and $b = 0.70$; 
			$a = 0.25$ and $b = 0.00$,
			the left and right truncation (or censored) thresholds 
			$d$ and $u$, respectively, 
			to the nearest two decimal places are chosen specifically 
			as $a = F(d)$ or $d = F^{-1}(a)$ and $1-b = F(u)$ 
			or $u = F^{-1}(1-b)$.
		\end{enumerate}
	\end{enumerate}
	
	
	\begin{sidewaystable}[btp!]
		\begin{center}
			\caption[Finite-sample performance evaluation of different Estimators 
			for exponential distribution.]{Finite-sample performance evaluation 
				of different Estimators for $Exp(\theta = 10)$.}
			\label{table:expSimStudy}
			{
				\begin{tabular}{ccc|ccc|ccc|ccc}
					\cline{1-12}
					& & & & & & & & & & & \\[-2.25ex]
					{} & \multicolumn{1}{c}{\multirow{2}{*}{$d_{(a)}$}} & 
					\multicolumn{1}{c|}{\multirow{2}{*}{$u_{(b)}$}} &
					MTuM & MCM & MTCM & MTuM & MCM & MTCM & MTuM & MCM & MTCM \\
					\cline{4-12}
					& & & & & & & & & & & \\[-2.25ex]
					{} & {} & {} & 
					\multicolumn{3}{c|}{$n=50$} &
					\multicolumn{3}{c|}{$n=100$} & 
					\multicolumn{3}{c}{$n=250$} \\
					\cline{1-12}
					& & & & & & & & & & & \\[-2.25ex]
					\multicolumn{1}{c}{\multirow{7}{*}{$\widehat{\theta}/\theta$}} 
					& $0.00_{(.00)}$ & $\infty_{(.00)}$ &
					1.00{\tiny (.000)} & 1.00{\tiny (.000)} & 1.00{\tiny (.000)} & 
					1.00{\tiny (.000)} & 1.00{\tiny (.000)} & 1.00{\tiny (.000)} & 
					1.00{\tiny (.000)} & 1.00{\tiny (.000)} & 1.00{\tiny (.000)} \\
					{} & $0.51_{(.05)}$ & $29.96_{(.05)}$ &
					1.03{\tiny (.001)} & 1.01{\tiny (.000)} & 1.02{\tiny (.000)} & 
					1.02{\tiny (.001)} & 1.00{\tiny (.000)} & 1.00{\tiny (.000)} & 
					1.01{\tiny (.000)} & 1.00{\tiny (.000)} & 1.00{\tiny (.000)} \\
					{} & $1.05_{(.10)}$ & $23.03_{(.10)}$ &
					1.08{\tiny (.002)} & 1.01{\tiny (.000)} & 1.01{\tiny (.001)} & 
					1.04{\tiny (.001)} & 1.01{\tiny (.000)} & 1.01{\tiny (.000)} & 
					1.01{\tiny (.000)} & 1.00{\tiny (.000)} & 1.00{\tiny (.000)} \\
					{} & $1.63_{(.15)}$ & $18.97_{(.15)}$ &
					-- & 1.01{\tiny (.000)} & 1.02{\tiny (.001)} & 
					1.07{\tiny (.002)} & 1.01{\tiny (.000)} & 1.01{\tiny (.000)} & 
					1.02{\tiny (.001)} & 1.00{\tiny (.000)} & 1.00{\tiny (.000)} \\
					{} & $2.88_{(.25)}$ & $13.86_{(.25)}$ &
					-- & 1.02{\tiny (.001)} & 1.03{\tiny (.001)} & 
					-- & 1.01{\tiny (.000)} & 1.02{\tiny (.000)} & 
					-- & 1.00{\tiny (.000)} & 1.01{\tiny (.000)} \\
					{} & $1.05_{(.10)}$ & $03.57_{(.70)}$ &
					-- & 1.08{\tiny (.001)} & 1.16{\tiny (.002)} &
					-- & 1.04{\tiny (.001)} & 1.07{\tiny (.001)} & 
					-- & 1.01{\tiny (.000)} & 1.02{\tiny (.000)} \\
					{} & $2.88_{(.25)}$ & $\infty_{(.00)}$ &
					1.00{\tiny (.000)} & 1.00{\tiny (.000)} & 1.00{\tiny (.000)} &
					1.00{\tiny (.000)} & 1.00{\tiny (.000)} & 1.00{\tiny (.000)} & 
					1.00{\tiny (.000)} & 1.00{\tiny (.000)} & 1.00{\tiny (.000)} \\
					\cline{1-12}
					& & & & & & & & & & & \\[-2.25ex]
					\multicolumn{1}{c}{\multirow{7}{*}{\sc Re}} 
					& $0.00_{(.00)}$ & $\infty_{(.00)}$ &
					1.00{\tiny (.004)} & 1.00{\tiny (.004)} & 1.00{\tiny (.004)} &
					1.00{\tiny (.003)} & 1.00{\tiny (.003)} & 1.00{\tiny (.003)} & 
					1.01{\tiny (.006)} & 1.01{\tiny (.006)} & 1.01{\tiny (.006)} \\
					{} & $0.51_{(.05)}$ & $29.96_{(.05)}$ &
					0.33{\tiny (.003)} & 0.88{\tiny (.005)} & 0.83{\tiny (.005)} &
					0.38{\tiny (.002)} & 0.90{\tiny (.004)} & 0.85{\tiny (.003)} & 
					0.42{\tiny (.002)} & 0.92{\tiny (.005)} & 0.86{\tiny (.004)} \\
					{} & $1.05_{(.10)}$ & $23.03_{(.10)}$ &
					0.08{\tiny (.009)} & 0.80{\tiny (.005)} & 0.70{\tiny (.004)} & 
					0.18{\tiny (.002)} & 0.83{\tiny (.003)} & 0.72{\tiny (.003)} & 
					0.23{\tiny (.001)} & 0.85{\tiny (.004)} & 0.74{\tiny (.003)} \\
					{} & $1.63_{(.15)}$ & $18.97_{(.15)}$ &
					-- & 0.73{\tiny (.005)} & 0.57{\tiny (.004)} & 
					0.06{\tiny (.007)} & 0.76{\tiny (.003)} & 0.60{\tiny (.002)} & 
					0.12{\tiny (.001)} & 0.78{\tiny (.004)} & 0.63{\tiny (.002)} \\
					{} & $2.88_{(.25)}$ & $13.86_{(.25)}$ &
					-- & 0.61{\tiny (.005)} & 0.35{\tiny (.003)} &
					-- & 0.64{\tiny (.002)} & 0.39{\tiny (.002)} & 
					-- & 0.67{\tiny (.003)} & 0.42{\tiny (.002)} \\
					{} & $1.05_{(.10)}$ & $03.57_{(.70)}$ &
					-- & 0.14{\tiny (.001)} & 0.04{\tiny (.004)} & 
					-- & 0.20{\tiny (.001)} & 0.10{\tiny (.001)} & 
					-- & 0.23{\tiny (.002)} & 0.13{\tiny (.001)} \\
					{} & $2.88_{(.25)}$ & $\infty_{(.00)}$ &
					0.74{\tiny (.002)} & 0.99{\tiny (.004)} & 0.74{\tiny (.002)} & 
					0.74{\tiny (.003)} & 0.99{\tiny (.003)} & 0.74{\tiny (.003)} & 
					0.75{\tiny (.004)} & 1.00{\tiny (.006)} & 0.75{\tiny (.004)} \\
					\hline\hline
					& & & & & & & & & & & \\[-2.25ex]
					{} & \multicolumn{1}{c}{\multirow{2}{*}{$d_{(a)}$}} & 
					\multicolumn{1}{c|}{\multirow{2}{*}{$u_{(b)}$}} &
					MTuM & MCM & MTCM & MTuM & MCM & MTCM & 
					{MTuM} & {MCM} & {MTCM} \\
					\cline{4-12}
					& & & & & & & & & & & \\[-2.25ex]
					{} & {} & {} & \multicolumn{3}{c|}{$n=500$} & 
					\multicolumn{3}{c|}{$n=1000$} &
					\multicolumn{3}{c}{$n \rightarrow \infty$} \\
					\hline
					& & & & & & & & & & & \\[-2.25ex]
					\multicolumn{1}{c}{\multirow{7}{*}{$\widehat{\theta}/\theta$}} 
					& $0.00_{(.00)}$ & $\infty_{(.00)}$ &
					1.00{\tiny (.000)} & 1.00{\tiny (.000)} & 1.00{\tiny (.000)} & 
					1.00{\tiny (.000)} & 1.00{\tiny (.000)} & 1.00{\tiny (.000)} &
					1 & 1 & 1 \\
					{} & $0.51_{(.05)}$ & $29.96_{(.05)}$ &
					1.00{\tiny (.000)} & 1.00{\tiny (.000)} & 1.00{\tiny (.000)} & 
					1.00{\tiny (.000)} & 1.00{\tiny (.000)} & 1.00{\tiny (.000)} &
					1 & 1 & 1 \\
					{} & $1.05_{(.10)}$ & $23.03_{(.10)}$ &
					1.01{\tiny (.000)} & 1.00{\tiny (.000)} & 1.00{\tiny (.000)} & 
					1.00{\tiny (.000)} & 1.00{\tiny (.000)} & 1.00{\tiny (.000)} & 
					1 & 1 & 1 \\
					{} & $1.63_{(.15)}$ & $18.97_{(.15)}$ &
					1.01{\tiny (.000)} & 1.00{\tiny (.000)} & 1.00{\tiny (.000)} & 
					1.01{\tiny (.000)} & 1.00{\tiny (.000)} & 1.00{\tiny (.000)} &
					1 & 1 & 1 \\
					{} & $2.88_{(.25)}$ & $13.86_{(.25)}$ &
					1.05{\tiny (.001)} & 1.00{\tiny (.000)} & 1.00{\tiny (.000)} & 
					1.02{\tiny (.000)} & 1.00{\tiny (.000)} & 1.00{\tiny (.000)} & 
					1 & 1 & 1 \\
					{} & $1.05_{(.10)}$ & $03.57_{(.70)}$ &
					-- & 1.01{\tiny (.000)} & 1.01{\tiny (.000)} & 
					-- & 1.00{\tiny (.000)} & 1.01{\tiny (.000)} & 
					1 & 1 & 1 \\
					{} & $2.88_{(.25)}$ & $\infty_{(.00)}$ &
					1.00{\tiny (.000)} & 1.00{\tiny (.000)} & 1.00{\tiny (.000)} & 
					1.00{\tiny (.000)} & 1.00{\tiny (.000)} & 1.00{\tiny (.000)} & 
					1 & 1 & 1 \\
					\hline
					& & & & & & & & & & & \\[-2.25ex]
					\multicolumn{1}{c}{\multirow{7}{*}{\sc Re}} 
					& $0.00_{(.00)}$ & $\infty_{(.00)}$ &
					1.00{\tiny (.001)} & 1.00{\tiny (.001)} & 1.00{\tiny (.000)} & 
					0.99{\tiny (.004)} & 0.99{\tiny (.004)} & 0.99{\tiny (.004)} & 
					1 & 1 & 1 \\
					{} & $0.51_{(.05)}$ & $29.96_{(.05)}$ &
					0.43{\tiny (.001)} & 0.91{\tiny (.001)} & 0.86{\tiny (.000)} & 
					0.44{\tiny (.002)} & 0.91{\tiny (.004)} & 0.86{\tiny (.003)} & 
					.442 & .918 & .868 \\
					{} & $1.05_{(.10)}$ & $23.03_{(.10)}$ &
					0.24{\tiny (.001)} & 0.84{\tiny (.001)} & 0.74{\tiny (.000)} & 
					0.25{\tiny (.001)} & 0.84{\tiny (.004)} & 0.74{\tiny (.003)} & 
					.257 &.848 & .749 \\
					{} & $1.63_{(.15)}$ & $18.97_{(.15)}$ &
					0.14{\tiny (.001)} & 0.78{\tiny (.001)} & 0.63{\tiny (.000)} & 
					0.14{\tiny (.001)} & 0.78{\tiny (.003)} & 0.63{\tiny (.002)} & 
					.152 & .787 & .638 \\
					{} & $2.88_{(.25)}$ & $13.86_{(.25)}$ &
					0.03{\tiny (.001)} & 0.67{\tiny (.001)} & 0.42{\tiny (.000)} & 
					0.04{\tiny (.000)} & 0.67{\tiny (.002)} & 0.43{\tiny (.001)} & 
					.047 & .679 & .433 \\
					{} & $1.05_{(.10)}$ & $03.57_{(.70)}$ &
					-- & 0.24{\tiny (.001)} & 0.14{\tiny (.000)} & 
					-- & 0.24{\tiny (.001)} & 0.15{\tiny (.001)} & 
					.001 & .250 & .156 \\
					{} & $2.88_{(.25)}$ & $\infty_{(.00)}$ &
					0.75{\tiny (.001)} & 0.99{\tiny (.001)} & 0.75{\tiny (.000)} &
					0.75{\tiny (.003)} & 0.99{\tiny (.004)} & 0.75{\tiny (.003)} & 
					.750 & .995 & .750 \\
					\hline\hline 
				\end{tabular}}
			\end{center}
		\end{sidewaystable}
		
		Simulation results are recorded in Table \ref{table:expSimStudy}. 
		The entries are mean values (with standard errors in parentheses)
		based on 100,000 samples.  
		The columns corresponding to 
		$n \rightarrow \infty$,
		represent analytic 
		$
		\mbox{ARE}
		\left(
		\widehat{\theta}_{\mbox{\tiny $\mathcal{C}$}}, 
		\widehat{\theta}_{\mbox{\tiny MLE}}
		\right)
		$ results with 
		$\mathcal{C} \in \{\mbox{MTuM, MCM, MTCM}\}$ and are found in
		Section \ref{sec:MTuM_ExpPareto}, not from simulations.
		Among these three columns, the first one is 
		$\mathcal{C} = MTuM$, second 
		$\mathcal{C} = \mbox{MCM}$,
		and third
		$\mathcal{C} = \mbox{MTCM}$. 
		As seen from Table \ref{table:expSimStudy},
		the ratio  
		$\widehat{\theta}/\theta$
		\label{mis:Mean2}
		of the exponential $\theta$ estimators 
		converges to the true asymptotic value of  
		$1$ very fast.
		Besides MTuM approach, the bias of all other procedures 
		disappears as soon as $n \ge 500$ and the estimators' RE's
		practically reach their ARE levels just for $n \ge 250$.
		Some of the finite sample entries for MTuM columns
		in Table \ref{table:expSimStudy} are not 
		reported, specially for the pair $(d,u) = (1.05,3.57)$,
		because the corresponding threshold pair
		$(d,u)$ does not satisfy the necessary 
		condition of Theorem \ref{thm:ExpTruncatedMean}
		for at least one generated sample. 
		It is evident from the entries that if the difference between
		the thresholds (i.e., $d$ and $u$) is smaller then the estimators 
		converge slower to the true values.
		Overall, as expected MCM perform the best in terms of balancing 
		between efficiency and robustness. 
		MTuM performs very poorly specially for small sample sizes in terms of
		efficiency but this approach produces highly robust estimators. 
		
		\section{Concluding Remarks}
		\label{sec:Conclusion}
		
		In this paper, we have developed the methods of 
		{\em truncated\/} (called MTuM),
		{\em censored\/} (called MCM), and 
		{\em insurance payment-type\/}
		(called MTCM) moments estimators for 
		completely observed ground-up 
		loss severity data.
		A series of theoretical results about estimators'
		existence and asymptotic normality are established. 
		Our analysis has established new connections between data truncation, 
		trimming, and censoring, which paves the way for more effective 
		modeling of non-linearly transformed loss data. 
		Further, as seen from Table 
		\ref{table:CombinedAREs1},
		there is clear trade-offs between efficiency and 
		robustness between newly designed estimators and 
		the corresponding MLEs when sample size is large.
		The finite sample performance, for various sample
		sizes, of all the estimators developed in this 
		paper has been investigated in detail for 
		single parameter Pareto model via
		simulation study.
		
		The results of this paper motivate open problems 
		and generate several ideas for further research. 
		First, most of the results of Section \ref{sec:MTuM_ExpPareto} 
		(beside Theorem \ref{thm:ifmtmmcm}) are limited to 
		complete exponentially (equivalently single parameter
		Pareto) distributed data but they could be extended 
		to more general situations and models.
		For example, similar estimation approaches 
		could be designed for (log) 
		location-scale and exponential dispersion families
		which could lead to more challenging non-linear equations 
		to be solved (see Theorems \ref{thm:ExpTruncatedMean},
		\ref{thm:MCMExistenceTheorem}, and \ref{thm:MTCMExistenceTheorem}). 
		Second, several contaminated loss severity models are proposed 
		in the literature 
		\citep[see, e.g.,][]{MR3474025,  MR3807461, MR2969968},
		so it could even produce a better model still maintaining 
		a reasonable balance between efficiency and robustness
		if one implements the procedures developed in this paper
		on the body part of the data and some heavier distributions 
		(say, for e.g., Pareto) on the right tail.
		Further, it is yet to measure how the newly designed estimation 
		procedures act with different risk analysis in practice. 
		
		{\baselineskip 5mm
			\renewcommand{\bibname}{References}
			\bibliography{ArXiv}
			\thispagestyle{plain}
			\pagestyle{plain} 
			\thispagestyle{plain}
		}
		
		\begin{appendices}
			
			\section{All four possible scenarios for Section 
				\ref{sec:mtum_asymptotic_properties}}
			\label{apdx:allScenarios}
			
			\textbf{Scenario 1}: $d_{j} \leq d_{j'} < u_{j} \leq u_{j'}$ 
			\setlength{\unitlength}{1cm}
			\thicklines
			\begin{picture}(1,-5)
			\put(2,0.1){\vector(1,0){8}}
			\put(2,0.1){\vector(-1,0){1}}
			\put(3,.1){\line(0,1){.2}}
			\put(5,.1){\line(0,1){.2}}
			\put(6,.1){\line(0,1){.2}}
			\put(9,.1){\line(0,1){.2}}
			\put(2.9,-0.3){$d_{j}$}
			\put(4.9,-0.3){$d_{j'}$}
			\put(5.9,-0.3){$u_{j}$}
			\put(8.9,-0.3){$u_{j'}$}
			\end{picture} 
			\begin{align*}
				Y_{jj'} & = h_{jj'}(X)\ID\{d_{jj'}<X \leq u_{jj'}\} 
				= 
				h_{jj'}(X)\ID\{d_{j'}<X \leq u_{j}\}, \\
				W_{jj'} & = Z_{j}\ID\{d_{j'}<X \leq u_{j}\}, 
				\quad \text{and} \quad 
				W_{j'j} = Z_{j'}\ID\{d_{j'}<X \leq u_{j}\}.
			\end{align*}
			\textbf{Scenario 2}: $d_{j} \leq d_{j'} < u_{j'} \leq u_{j}$ 
			\setlength{\unitlength}{1cm}
			\thicklines
			\begin{picture}(1,-5)
			\put(2,0.1){\vector(1,0){8}}
			\put(2,0.1){\vector(-1,0){1}}
			\put(3,.1){\line(0,1){.2}}
			\put(5,.1){\line(0,1){.2}}
			\put(6,.1){\line(0,1){.2}}
			\put(9,.1){\line(0,1){.2}}
			\put(2.9,-0.3){$d_{j}$}
			\put(4.9,-0.3){$d_{j'}$}
			\put(5.9,-0.3){$u_{j'}$}
			\put(8.9,-0.3){$u_{j}$}
			\end{picture} 
			\begin{align*}
				Y_{jj'} & = h_{jj'}(X)\ID\{d_{jj'}<X \leq u_{jj'}\} 
				= 
				h_{jj'}(X)\ID\{d_{j'}<X \leq u_{j'}\}, \\
				W_{jj'} & = Z_{j}\ID\{d_{j'}<X \leq u_{j'}\}, \quad \text{and} \quad 
				W_{j'j} = Z_{j'}\ID\{d_{j'}<X \leq u_{j'}\}.
			\end{align*}
			\textbf{Scenario 3}: $d_{j'} \leq d_{j} < u_{j} \leq u_{j'}$ 
			\setlength{\unitlength}{1cm}
			\thicklines
			\begin{picture}(1,-5)
			\put(2,0.1){\vector(1,0){8}}
			\put(2,0.1){\vector(-1,0){1}}
			\put(3,.1){\line(0,1){.2}}
			\put(5,.1){\line(0,1){.2}}
			\put(6,.1){\line(0,1){.2}}
			\put(9,.1){\line(0,1){.2}}
			\put(2.9,-0.3){$d_{j'}$}
			\put(4.9,-0.3){$d_{j}$}
			\put(5.9,-0.3){$u_{j}$}
			\put(8.9,-0.3){$u_{j'}$}
			\end{picture} 
			\begin{align*}
				Y_{jj'} 
				& = 
				h_{jj'}(X)\ID\{d_{jj'}<X \leq u_{jj'}\} 
				= 
				h_{jj'}(X)\ID\{d_{j}<X \leq u_{j}\}, \\
				W_{jj'} 
				& = 
				Z_{j}\ID\{d_{j}<X \leq u_{j}\}, \quad \text{and} \quad
				W_{j'j} 
				= 
				Z_{j'}\ID\{d_{j}<X \leq u_{j}\}.
			\end{align*}
			\textbf{Scenario 4}: $d_{j'} \leq d_{j} < u_{j'} \leq u_{j}$ 
			\setlength{\unitlength}{1cm}
			\thicklines
			\begin{picture}(1,-5)
			\put(2,0.1){\vector(1,0){8}}
			\put(2,0.1){\vector(-1,0){1}}
			\put(3,.1){\line(0,1){.2}}
			\put(5,.1){\line(0,1){.2}}
			\put(6,.1){\line(0,1){.2}}
			\put(9,.1){\line(0,1){.2}}
			\put(2.9,-0.3){$d_{j'}$}
			\put(4.9,-0.3){$d_{j}$}
			\put(5.9,-0.3){$u_{j'}$}
			\put(8.9,-0.3){$u_{j}$}
			\end{picture} 
			\begin{align*}
				Y_{jj'} 
				& = 
				h_{jj'}(X)\ID\{d_{jj'}<X \leq u_{jj'}\}
				= 
				h_{jj'}(X)\ID\{d_{j}<X \leq u_{j'}\}, \\
				W_{jj'} & = Z_{j}\ID\{d_{j}<X \leq u_{j'}\}, 
				\quad \text{and} \quad
				W_{j'j} = Z_{j'}\ID\{d_{j}<X \leq u_{j'}\}.
			\end{align*}
			Therefore, depending on the scenario the expected values are given by:
			\begin{align*}
				\mu_{Y_{jj'}} & = \E[Y_{jj'}] & \mu_{W_{jj'}} 
				& = 
				\E[W_{jj'}] \\
				& =
				\int_{F(d_{jj'}|\bm{\theta})}^{F(u_{jj'}|\bm{\theta})}h_{jj'}
				\left(F^{-1}(v|\bm{\theta})\right) dv, & {} 
				& = 
				\int_{F(d_{jj'}|\bm{\theta})}^{F(u_{jj'}|\bm{\theta})}h_{j}
				\left(F^{-1}(v|\bm{\theta})\right) dv. 
			\end{align*}
			
			\vspace{5mm}
			
			\section{Proofs}
			\label{apdx:proofs}
			
			\textbf{\textit{Proof of Theorem \ref{thm:MTuMMuAsyDist}:}}
			\begin{proof}[\unskip\nopunct]
				Clearly, $g_{\bm{V}}(\bm{\mu}_{\bm{V}}) = \left(\frac{\mu_{Y_{1}}}{p_{1}},\ldots,\frac{\mu_{Y_{k}}}{p_{k}} \right)=:\left(\mu_{1},\ldots,\mu_{k}\right) 
				=: 
				\bm{\mu}$.
				From Lemma \ref{lemma:differentialFun}, it follows that 
				$$\bm{D}_{\bm{V}} := \left[\left. \frac{\partial g_{j}}{\partial x_{j'}}\right\vert_{\bm{x}=\bm{\mu}_{\bm{V}}}\right]_{k\times 2k}
				= 
				\left[d_{\bm{V},jj'}\right]_{k\times 2k},$$
				where 
				\begin{equation*}
					d_{\bm{V},jj'} 
					:= 
					\begin{dcases}
						\frac{1}{p_{j'}}, & \ \ \ \ \ \text{if } 1 \leq j = j' \leq{k}; 
						\\
						-\frac{\mu_{Y_{j}}}{p_{j}^{2}}, & \ \ \ \ \ \text{if } j'-j = k; 
						\\
						0,  & \ \ \ \ \ \text{otherwise.}
					\end{dcases}
				\end{equation*}
				
				Now, with an application of delta method corresponding with
				the function $g_{\bm{V}}$ above, \citep[see][\S3.3 Theorem A]{MR595165}, 
				we have 
				\[
				\left(\widehat{\mu}_{1},\ldots,\widehat{\mu}_{k}\right) \sim \mathcal{AN}\left(g_{\bm{V}}(\bm{\mu}_{\bm{V}}) = \bm{\mu},\frac{1}{n}\bm{D_{\bm{V}}\Sigma_{\bm{V}}D_{\bm{V}}'}\right). \qedhere
				\]
			\end{proof}
			
			\noindent
			\textbf{\textit{Proof of Theorem \ref{thm:theTruncatedMeanVariance}:}}
			\begin{proof}[\unskip\nopunct]
				The r.v. $Y$ can be expressed in the form of 
				$$Y = X\wedge u - u\ID \{u < X <\infty\} - X\wedge d + d\ID \{d < X <\infty\}.$$
				Define, $I_{a,b} := \ID \{a < X <b\}$. Therefore,
				\begin{align*}
					\mu_{Y} 
					& = 
					\E[Y] 
					= 
					\E[X\wedge u] - \E[uI_{u,\infty}] - \E[X\wedge d] + \E[dI_{d,\infty}] \\
					& = \theta (1-e^{-\frac{u}{\theta}}) - u e^{-\frac{u}{\theta}} 
					- \theta (1-e^{-\frac{d}{\theta}}) + d e^{-\frac{d}{\theta}} =
					\theta \left(e^{-\frac{d}{\theta}} - e^{-\frac{u}{\theta}}\right)
					+ de^{-\frac{d}{\theta}} - ue^{-\frac{u}{\theta}}.
				\end{align*}
				Since $Y = X\wedge u - u\ID \{u < X <\infty\} 
				- X\wedge d + d\ID \{d < X <\infty\},$
				then
				\begin{eqnarray*}
					Y^{2} 
					& = & 
					(X\wedge u)^{2} - (X\wedge d)^{2} -2d\left[X\wedge u
					- X\wedge d\right] - u^{2}I_{u,\infty} - d^{2}I_{d,\infty} 
					+ 2d\left(XI_{d,u}+uI_{u,\infty}\right).
				\end{eqnarray*}
				Therefore, for $X \sim Exp(\theta)$ 
				then $\mu_{Y^{2}} := \E[Y^{2}]$ 
				is computed as below: 
				\begin{eqnarray*}
					\mu_{Y^{2}} 
					& = & 
					\E[Y^{2}] \\
					& = & 
					\E[(X\wedge u)^{2}] - \E[(X\wedge d)^{2}] 
					- 2d\left[\E[X\wedge u] - \E[X\wedge d]\right] 
					- u^{2}\E[I_{u,\infty}] \\
					& & {} - d^{2}\E[I_{d,\infty}] + 2d\left(\E[XI_{d,u}]
					+ u\E[I_{u,\infty}]\right) \\
					& = & 
					2\theta^{2}\Gamma\left(3;\frac{u}{\theta}\right)
					+ u^{2}e^{-\frac{u}{\theta}} 
					- 2\theta^{2}\Gamma\left(3;\frac{d}{\theta}\right) 
					- d^{2}e^{-\frac{d}{\theta}} \\
					& & {} - 2d\left[\theta\left(1-e^{-\frac{u}{\theta}}\right)
					-\theta\left(1-e^{-\frac{d}{\theta}}\right) \right] 
					- u^{2}e^{-\frac{u}{\theta}} -d^{2}e^{-\frac{d}{\theta}}	\\
					& & {} 
					+ 2d\left[\theta e^{-\frac{d}{\theta}}+de^{-\frac{d}{\theta}}-\theta e^{-\frac{u}{\theta}}-ue^{-\frac{u}{\theta}}+ue^{-\frac{R}{\theta}}\right] \\	
					& = & 
					2\theta^{2}\left(\Gamma\left(3;\frac{u}{\theta}\right) 
					- \Gamma\left(3;\frac{d}{\theta}\right)\right).
				\end{eqnarray*}
				Therefore,
				\[
				\sigma_{Y}^{2} 
				= 
				\mu_{Y^{2}} - \mu_{Y}^{2} 
				=  
				2\theta^{2}\left(\Gamma\left(3;\frac{u}{\theta}\right) 
				- \Gamma\left(3;\frac{d}{\theta}\right)\right) - \mu_{Y}^{2}. 
				\qedhere
				\]
			\end{proof}
			
			\noindent
			\textbf{\textit{Proof of Theorem \ref{thm:gtTPI}:}}
			\begin{proof}[\unskip\nopunct]
				Note that the parameter vector is given by 
				$\bm{\theta} 
				= 
				(\alpha,x_{0})$ with $x_{0}$ 
				known in advance.
				The population version of $\widehat{\mu}_{\mbox{\tiny MTuM}}$ 
				is given by 
				\begin{eqnarray*}
					\mu_{\mbox{\tiny MTuM}}
					& = & 
					\E[h(Y) | d < Y \leq u] 
					= 
					\frac{\E[h(Y)\ID \{d < Y \leq u\}]}{F(d|\bm{\theta})-F(d|\bm{\theta})} 
					=
					\frac{\int_{d}^{u}h(y)f(y|\bm{\theta}) \, dy}{F(u|\bm{\theta})-F(d|\bm{\theta})} \\
					& = & 
					\frac{\int_{F(d|\bm{\theta})}^{F(u|\bm{\theta})}h(F^{-1}(v|\bm{\theta})) \, dv}{F(u|\bm{\theta})-F(d|\bm{\theta})} 
					=
					-\frac{\int_{F(d|\bm{\theta})}^{F(u|\bm{\theta})}\log{(1-v)} \, dv}
					{\alpha (F(u|\bm{\theta}) - F(d|\bm{\theta}))} \\
					& = & -\frac{1}{\alpha\left(\left(\frac{x_{0}}{d}\right)^{\alpha} 
						- \left(\frac{x_{0}}{u}\right)^{\alpha} \right)} \\
					& & {} \times \left[F(d|\bm{\theta})
					-F(u|\bm{\theta})+\alpha(1-F(d|\bm{\theta}))\log{\left(\frac{x_{0}}{d}\right)} - \alpha(1-F(u|\bm{\theta}))\log{\left(\frac{x_{0}}{u}\right)}\right] \\
					& = & -\frac{1}{\alpha\left(\left(\frac{x_{0}}{d}\right)^{\alpha} 
						- \left(\frac{x_{0}}{u}\right)^{\alpha} \right)}  
					\left[\left(\frac{x_{0}}{u}\right)^{\alpha} 
					- \left(\frac{x_{0}}{d}\right)^{\alpha} + \alpha \left(\frac{x_{0}}{d}\right)^{\alpha}\log{\left(\frac{x_{0}}{d}\right)} 
					-\alpha \left(\frac{x_{0}}{u}\right)^{\alpha}
					\log{\left(\frac{x_{0}}{u}\right)} \right] \\
					& = & \frac{x_{0}^{\alpha} (du)^{\alpha}}{\alpha x_{0}^{\alpha} (du)^{\alpha}
						(u^{\alpha} - d^{\alpha})}\left[u^{\alpha}
					\left(1-\alpha \log{\left(\frac{x_{0}}{d}\right)}\right) 
					- d^{\alpha}\left(1-\alpha \log{\left(\frac{x_{0}}{u}\right)}\right)\right] \\
					& = & \frac{1}{\alpha (u^{\alpha} - d^{\alpha})}\left[u^{\alpha}\left(1-\alpha \log{\left(\frac{x_{0}}{d}\right)}\right) - d^{\alpha}\left(1-\alpha \log{\left(\frac{x_{0}}{u}\right)}\right)\right] \\
					& = & \frac{A_{du}}{\alpha (u^{\alpha} - d^{\alpha})}.
				\end{eqnarray*}		
				Now, to establish the proof of the statement, it is 
				enough to prove that the function $g_{du}$ is strictly decreasing 
				with respect to $\alpha$. For that,
				\begin{align*}
					g_{du}'(\alpha) & = 	\frac{dg_{du}(\alpha)}{d\alpha} 
					= 
					\frac{(du)^{\alpha}\alpha^{2}\left(\log\left(\frac{u}{d}\right)\right)^{2} 
						- (u^{\alpha} - d^{\alpha})^{2}}{\alpha^{2}(u^{\alpha} - d^{\alpha})^{2}}.
				\end{align*} 
				Now, in order to show that $g_{du}^{'}(\alpha) < 0$, it is enough to establish $(du)^{\alpha}\alpha^{2}\left(\log\left(\frac{u}{d}\right)\right)^{2} 
				- (u^{\alpha} - d^{\alpha})^{2} < 0$ which is equivalent to establish that $(du)^{\frac{\alpha}{2}}\alpha\log\left(\frac{u}{d}\right) < u^{\alpha} 
				- d^{\alpha}$. 
				Now,
				\begin{align*}
					{} & (du)^{\frac{\alpha}{2}}\alpha\log\left(\frac{u}{d}\right) 
					< u^{\alpha} - d^{\alpha} \\
					\iff \quad & \alpha\log\left(\frac{u}{d}\right) 
					< \left(\frac{u}{d}\right)^{\frac{\alpha}{2}} 
					- \left(\frac{u}{d}\right)^{-\frac{\alpha}{2}} \\
					\iff \quad & \alpha\log\left(\frac{u}{d}\right) 
					< 2\sinh{\left(\frac{\alpha}{2} \log\left(\frac{u}{d}\right)\right)} \\
					\iff \quad & \frac{\alpha}{2}\log\left(\frac{u}{d}\right) 
					< \sinh{\left(\frac{\alpha}{2} \log\left(\frac{u}{d}\right)\right)}.
				\end{align*} 
				But we know that $x<\sinh{x}$ for all $x>0$, therefore, 
				$g_{du}^{'}(\alpha) < 0$ for all $\alpha>0$ 
				which implies that $g_{du}$ is strictly decreasing. 
				Finally, note that
				\begin{align*}
					\lim_{\alpha \rightarrow 0+} g_{du}(\alpha) 
					& = 
					\frac{{(\log(u))^{2}}-(\log(d))^{2}-2\log(u)\log\left(\frac{x_{0}}{d}\right)
						+ 2\log(d)\log\left(\frac{x_{0}}{u}\right)}{2\log\left(\frac{u}{d}\right)}, 
					\ \mbox{and} \\
					\lim_{\alpha \rightarrow \infty} g_{du}(\alpha) 
					& = 
					-\log{\left(\frac{x_{0}}{d}\right)}. \qedhere
				\end{align*}
			\end{proof}
			
		\end{appendices}
		
	\end{document}